\documentclass[a4paper,12pt]{article}
\pdfoutput=1
\usepackage[margin=1in]{geometry}
\usepackage{graphicx,xcolor}
\usepackage{bbold}
\usepackage{mathtools,braket,blkarray}
\allowdisplaybreaks
\usepackage{amssymb,amsfonts,bbm,relsize}
\usepackage[width=0.9\textwidth,footnotesize]{caption}
\usepackage[font=scriptsize]{subcaption}
\usepackage{cite,hyperref,url}
\hypersetup{pageanchor=false}
\usepackage{booktabs,multirow,makecell}
\usepackage[utf8]{inputenc}
\usepackage{cleveref}

\newcommand{\overbar}[1]{\mkern 1.5mu\overline{\mkern-1.5mu#1\mkern-1.5mu}\mkern 1.5mu}
\newcommand{\pmatr}[1]{\begin{pmatrix} #1 \end{pmatrix}}

\begin{document}

\begin{titlepage}
\begin{center}
{\bf\Large   $\mathbf{SO(10)}\times  \mathbf{S_4} $  Grand Unified Theory of Flavour
\\ and Leptogenesis   } \\[12mm]
Francisco~J.~de~Anda$^{\dagger}$%
\footnote{E-mail: \texttt{franciscojosedea@gmail.com}},
Stephen~F.~King$^{\star}$%
\footnote{E-mail: \texttt{king@soton.ac.uk}},
Elena~Perdomo$^{\star}$%
\footnote{E-mail: \texttt{e.perdomo-mendez@soton.ac.uk}}
\\[-2mm]

\end{center}
\vspace*{0.50cm}
\centerline{$^{\star}$ \it
School of Physics and Astronomy, University of Southampton,}
\centerline{\it
SO17 1BJ Southampton, United Kingdom }
\vspace*{0.2cm}
\centerline{$^{\dagger}$ \it
Tepatitl{\'a}n's Institute for Theoretical Studies, C.P. 47600, Jalisco, M{\'e}xico}
\vspace*{1.20cm}

\begin{abstract}
{\noindent
We propose a
Grand Unified Theory of Flavour, based on $SO(10)$ together with a non-Abelian discrete group $S_4$, 
under which the unified three quark and lepton 16-plets are unified into a single triplet $3'$.
The model involves a further discrete group $ \mathbb{Z}_4^R\times \mathbb{Z}_4^3$ 
which controls the Higgs and flavon symmetry breaking sectors.
The CSD2 flavon vacuum alignment is discussed, along with the GUT breaking potential and
the doublet-triplet splitting, and proton decay is shown to be under control. 
The Yukawa matrices are derived in detail, from renormalisable diagrams, 
and neutrino masses emerge from the type I seesaw mechanism.
A full numerical fit is performed with 15 input parameters generating 19 presently constrained observables,
taking into account supersymmetry threshold corrections.
The model predicts a normal neutrino mass ordering with a CP oscillation phase
of $260^{\circ}$, an atmospheric angle in the first octant and neutrinoless double beta 
decay with $m_{\beta \beta}= 11$ meV.
We discuss $N_2$ leptogenesis, which fixes the second right-handed neutrino mass to be
$M_2\simeq 2\times 10^{11}$ GeV,
in the natural range predicted by the model.
}
\end{abstract}
\end{titlepage}

\section*{Introduction}

The Standard Model (SM)~\cite{Olive:2016xmw}, though highly successful, does not address the origin of neutrino mass and lepton mixing \cite{Fukuda:1998mi}. One attractive possibility is the type I seesaw mechanism, which can account for the smallness of neutrino masses
by introducing three right-handed neutrinos with very large Majorana masses \cite{Minkowski:1977sc}. Such right-handed neutrinos
arise very naturally from $SO(10)$ Grand Unified Theories (GUTs) \cite{Fritzsch} in which a single family of quarks and leptons,
together with a right-handed neutrino, is unified into a single 16-plet. Supersymmetry (SUSY) is then 
naturally suggested for gauge coupling unification and to ameliorate the gauge hierarchy problem. 
However the origin of the three families,
and their hierarchical masses are not explained by traditional $SO(10)$ SUSY GUTs. 

The almost tri-bimaximal lepton mixing  
observed over recent years \cite{nobel}, combined with a reactor angle 
of order $8.5^{\circ}$ \cite{King:2012vj}, suggests that some sort of non-Abelian family symmetry
may be at work in the lepton sector \cite{King:2017guk}.
The first models to consider a non-Abelian $SU(3)$ symmetry as an explanation of bi-large lepton 
mixing were put forward in \cite{King:2001uz}.
Models based on $SO(10)$ with a non-Abelian discrete symmetry were first proposed in
 \cite{so10, Bjorkeroth:2015uou}, 
and further flavoured GUTs were considered in \cite{nonAbelian}. A more general study of flavour symmetries in $SO(10)$ can be found in \cite{Ivanov:2015xss}.
Here we shall be interested in a SUSY GUT 
theory of flavour in which all quarks and leptons are fitted into a single $(3,16)$ representation of 
$S_4\times SO(10)$ \cite{LeeMohapatra,Bjorkeroth:2017ybg}.
While the former model predicted a zero reactor angle \cite{LeeMohapatra}, the latter model
\cite{Bjorkeroth:2017ybg} was based on 
CSD3~\footnote{CSD refers to ``constrained sequential dominance'' first introduced in \cite{King:2005bj}.
In this paper CSD is simply used as a label which refers to a particular flavon vacuum alignment as discussed later.
Such vacuum alignments motivates the choice of $S_4$ as the family symmetry,
as discussed by Luhn {\it et al} \cite{King:2013iva}.}
flavon vacuum alignment \cite{King:2013iva}, leading to approximate tri-bimaximal mixing with 
the correct value of the reactor angle.
However the latter model is so far incomplete since it did not include any explicit discussion of the flavon vacuum alignment,
or GUT breaking potential, and also did not include any discussion of leptogenesis. 

In the present paper we consider a more complete $S_4\times SO(10)$ SUSY GUT of flavour,
which also involves a further discrete group $ \mathbb{Z}_4^R\times \mathbb{Z}_4^3$ 
which controls the Higgs and flavon symmetry breaking sectors.
In the model here, we prefer the simpler CSD2~\cite{Antusch:2011ic} vacuum alignment,
which, in conjunction with small charged lepton corrections arising from the $SO(10)$ structure of Yukawa matrices, 
is capable of yielding 
the desired reactor angle. It also allows successful leptogenesis, as discussed below.
Here the flavon vacuum alignment potential is discussed, along with the GUT breaking potential and
the doublet-triplet splitting, and proton decay are shown to be under control. 
The Yukawa matrices are derived in detail, from renormalisable diagrams, 
and neutrino masses emerge from the type I seesaw mechanism.
A full numerical fit is performed with 15 input parameters describing 19 observables,
taking into account supersymmetry threshold corrections.
The model predicts a normal neutrino mass ordering with a CP oscillation phase
of $260^{\circ}$, an atmospheric angle in the first octant and neutrinoless double beta 
decay with $m_{\beta \beta}= 11$ meV.
We also discuss $N_2$ leptogenesis \cite{DiBari:2005st,DiBari:2015svd}, which fixes the second right-handed neutrino mass 
$M_2\simeq 2\times 10^{11}$ GeV,
in the natural range predicted by the model
\footnote{Interestingly we find that 
$N_2$ leptogenesis is not consistent with the earlier model based on CSD3 vacuum alignment \cite{Bjorkeroth:2017ybg},
which is a significant motivation for considering the new model based on CSD2.}.

The layout of the remainder of the paper is as follows. In Section~\ref{sec:overviewmodel} we describe the symmetries of the model and the superfields related to the low energy fields. In Section~\ref{sec:yuk} we list the complete set of fields together with the effective Yukawa terms they generate. Section~\ref{sec:vaal} shows how the flavon VEVs are aligned in the CSD2 direction. In Section~\ref{sec:sb} we show the symmetry breaking superpotential that produces a hierarchy between the flavon VEVs and drives them, together with the GUT breaking fields. Section~\ref{sec:dt} shows how doublet-triplet splitting is achieved. In Section~\ref{sec:pd}, proton decay is discussed. In Section~\ref{sec:yukd} we give the complete Yukawa superpotential and the fermion mass matrices structure arising from it. In Section~\ref{sec:fit} we give a numerical fit of model parameters to data, as well as the parametrization of SUSY threshold corrections and the parameter counting of the model. In Section~\ref{sec:lept} we show how the model can achieve successful $N_2$ leptogenesis. Section~\ref{sec:con} lists our conclusions.

\section{Overview of the model}
\label{sec:overviewmodel}
The symmetry of the model is $SO(10)\times S_4\times  \mathbb{Z}_4^R\times \mathbb{Z}_4^3$. The model has a gauge symmetry $SO(10)$ which is the GUT symmetry. The symmetry $S_4$ is the flavour symmetry which gives the specific CSD2 structure to the fermion mass matrices. The $\mathbb{Z}_4^R$ is an R symmetry while the other three $\mathbb{Z}_4$'s are shaping symmetries.  Furthermore, we assume that the GUT theory is invariant under trivial CP symmetry, which is spontaneously broken by the complex VEVs of the flavons.

\begin{table}[ht]
\begin{subtable}[t]{0.5\textwidth}
\centering
	\begin{tabular}[t]{| c | c@{\hskip 5pt}c | c c c c|}
\hline
\multirow{2}{*}{\rule{0pt}{4ex}Field}	& \multicolumn{6}{c |}{Representation} \\
\cline{2-7}
\rule{0pt}{3ex}			& $S_4$ & $SO(10)$ &  $ \mathbb{Z}^R_4$&$\mathbb{Z}_{4}$ &$\mathbb{Z}_{4}$ & $\mathbb{Z}_4$ \\ [0.75ex]
\hline \hline
\rule{0pt}{3ex}%
$\psi$ 			& $3^\prime$ & 16 & 1 &0 & 0& 0\\
\rule{0pt}{3ex}%
$H_{10}^{u}$ & 1 & 10 & $0$ & 0 &0&0 \\
$H_{10}^{d}$ & 1 & 10 & $0$ & 0 &2 &0\\
$H_{\overbar{16}}$ & 1 & $\overbar{16}$ & $0$ & 0 &0 & 0\\
$H_{16}$ & 1 & 16 & $0$ & 0 &1 &0 \\
$H_{45}^{X,Y}$ & 1 & 45 & 0 & 0&1 &0\\
$H_{45}^{W,Z}$ & 1 & 45 & $0$ & 2 &0 &0 \\
\rule{0pt}{3ex}%
$H_{45}^{B-L}$ & 1 & 45 & $2$ & 0 &2 &0\\
$\zeta$ & 1 &1 &0 &0 &2&0\\
\hline
\end{tabular}
\caption{Matter, and Higgs superfields.}
\end{subtable}
\begin{subtable}[t]{0.5\textwidth}
\centering
	\begin{tabular}[t]{| c | c@{\hskip 5pt}c | c c c c|}
\hline
\multirow{2}{*}{\rule{0pt}{4ex}Field}	& \multicolumn{6}{c |}{Representation} \\
\cline{2-7}
\rule{0pt}{3ex}			& $S_4$ & $SO(10)$ &  $ \mathbb{Z}^R_4$&$\mathbb{Z}_{4}$ &$\mathbb{Z}_{4}$ & $\mathbb{Z}_4$ \\ [0.75ex]
\hline \hline
\rule{0pt}{3ex}%
\rule{0pt}{3ex}%
$\phi_1$ & $3^\prime$ &1 &0 &2 &2&0\\
$\phi_2$ & $3^\prime$ &1 &0 & 2 &0&0\\
$\phi_3$ & $3^\prime$ &1 &0 & 0 &2&0 \\
\rule{0pt}{3ex}%
$\phi_{S,U}$ & $3^\prime$ &1 &0 & 0 &0&1 \\
$\phi_{T}$ & 3 &1 &0 & 1 &0& 1\\
$\xi $&1 & 1 & 0 & 3& 0& 2\\ 
$\phi_{t} $&3 & 1 & 0 & 0&1 & 3\\ 
\hline
\end{tabular}
\caption{Flavon superfields.}
\end{subtable}
\caption{Field content of the model that relates directly to the low energy fields.}
\label{tab:funfields}
\end{table}

In the table \ref{tab:funfields} we present the fields that contain the Higgs, flavons and matter fields, which are relevant to Yukawa sector.  The field $\psi$ contains the full SM fermion content. The fields $H^{u,d}_{10}$ contain the MSSM Higgs doublets $h_{u,d}$ respectively. The $H_{\overbar{16}}$ breaks $SO(10)\to SU(5)$ and gives masses to the right handed neutrinos (RHN). The $H_{45}$'s break $SU(5)\to SM$ and introduce the necessary Clebsch-Gordan (CG) relations to generate correct charged lepton and down quark masses. The flavon fields $\phi_i$, with $i=1,2,3$ break $S_4$ completely with the
CSD2 vacuum alignment~\cite{Antusch:2011ic},
\begin{equation}\begin{split}
\braket{\phi_1}= v_1\left(\begin{array}{c}1\\0\\2\end{array}\right),\ \ \braket{\phi_2}=v_2\left(\begin{array}{c}0\\1\\-1\end{array}\right),\ \ \braket{\phi_3}=v_3\left(\begin{array}{c}0\\1\\0\end{array}\right),
\label{eq:phfla}
\end{split}\end{equation}
with $|v_1|\ll |v_2|\ll |v_3|$. This CSD2 flavon alignment is fixed by a superpotential as discussed in Sec.~\ref{sec:vaal}.

With these fields, a very specific mass structure for the SM fermion fields is generated. For the up-type quark and the neutrino sectors, the Yukawa terms look like
\begin{equation}
H^u_{10} (\psi \phi_1)(\psi \phi_1)+ 
H^u_{10} (\psi \phi_2)(\psi \phi_2) + 
H^u_{10} (\psi \phi_3)(\psi \phi_3),
\label{eq:introup}
\end{equation}
where the brackets denote $S_4$ singlet contractions. Each of these terms generates
a rank-1 matrix. The hierarchy between the flavon VEVs, shown in Sec  \ref{sec:sb}, gives a natural explanation of the hierarchical Yukawa couplings 
	$y_u\sim v_1^2/M_\mathrm{GUT}^2$, 
	$y_c\sim v_2^2/M_\mathrm{GUT}^2$, 
	$y_t\sim v_3^2/M_\mathrm{GUT}^2$. The RHN Majorana masses are similar to Eq.~\ref{eq:introup} replacing $H^u_{10}$ by $H_{\overbar{16}}H_{\overbar{16}}$. The fact that the RHN masses have the same structure as the Dirac neutrino masses generate exactly the same structure for the left handed neutrino Majorana masses, as shown in Sec.~\ref{sec:seesaw}, after the seesaw mechanism. 

For the down-type quark and the charged lepton sectors, the Yukawa terms look like
\begin{equation}
H^d_{10} (\psi \phi_1)(\psi \phi_2) + 
H^d_{10} (\psi \phi_2)(\psi \phi_2) + 
H^d_{10} (\psi \phi_3)(\psi \phi_3) +
H^d_{10} (\psi \psi)_{3^\prime}(\phi_3),
\label{eq:introdown}
\end{equation}
where the brackets denote $S_4$ singlet contractions apart from the $3^\prime$ contraction which is necessary to combine with 
$\phi_3 \sim 3^\prime$ into a singlet. They have a different structure compared to the up sector, due to 
a mixing term between the flavons $\phi_1$ and $\phi_2$, which explains why there is a milder hierarchy in the down  and charged lepton sectors compared to the up one. It also introduces a texture zero in the (1,1) element of the down Yukawa matrix, reproducing the GST relation \cite{Gatto:1968ss}, i.e. the Cabibbo angle is predicted to be $\theta_{12}^q \simeq \sqrt{y_d/y_s}$. With this setup the full SM fermion masses are generated in a very specific and predictive way, this being the main aim of the paper.

After GUT symmetry breaking, all the messenger fields and adjoints obtain a GUT scale mass. Furthermore, the triplets inside the $H^{u,d}_{10}$ also get a GUT scale mass through the Dimopoulos-Wiclzeck mechanism \cite{DW}, as shown in the Sec.~\ref{sec:dt}. This way we make sure that at low energies, only the MSSM remains.

\section{Effective Yukawa structure}
\label{sec:yuk}

\begin{table}[ht]
\begin{subtable}[t]{0.5\textwidth}
\centering
	\begin{tabular}[t]{| c | c@{\hskip 5pt}c | c c c c|}
\hline
\multirow{2}{*}{\rule{0pt}{4ex}Field}	& \multicolumn{6}{c |}{Representation} \\
\cline{2-7}
\rule{0pt}{3ex}			& $S_4$ & $SO(10)$ &  $ \mathbb{Z}^R_4$&$\mathbb{Z}_{4}$ &$\mathbb{Z}_{4}$ & $\mathbb{Z}_4$ \\ [0.75ex]
\hline \hline
\rule{0pt}{3ex}%
$\bar{\chi}_1$ & 1 &$\overbar{16}$  &1 &$2$ &$2$&0\\
${\chi}_1$ & 1 &$16$& 1 & $0$&$2$&$0$\\
$\bar{\chi}_2$ & 1 &$\overbar{16}$  &1 & $2$&$0$&$0$\\
${\chi}_2$ & 1 &16 &1 &$0$ &$0$&$0$\\
$\bar{\chi}_3$ & 1 &$\overbar{16}$  &1 &$0$ &$2$&$0$ \\
$\chi_3$ & 1&$16$ &1 & $2$& $2$&$0$\\
\rule{0pt}{3ex}%
${\chi}_3^d$ & 1 &$16$& 1& $0$&$1$&$0$\\
${\chi}_2^d$ & 1 &16 &1 &$2$ &$3$&$0$\\
\hline
\rule{0pt}{3ex}%
$\bar{\chi}_u$ & 1 &$\overbar{16}$ &2 &$0$ &$0$&0\\
$\chi_u$ & 1 &$16$ &0 &$0$ &$2$&0\\
$\bar{\chi}_d$ & 1 &$\overbar{16}$ &0 &$0$ &$1$&0\\
$\chi_d$ & 1 &$16$ &2 &$0$ &$1$&0\\
$\zeta_1$ & 1 &$45$ &2 &$0$ &$3$&0\\
$\zeta_2$ & 1 &$45$ &0 &$0$ &$3$&0\\
\hline
\end{tabular}
\caption{Messenger superfields.}
\end{subtable}
\begin{subtable}[t]{0.5\textwidth}
\centering
	\begin{tabular}[t]{| c | c@{\hskip 5pt}c | c c c c|}
\hline
\multirow{2}{*}{\rule{0pt}{4ex}Field}	& \multicolumn{6}{c |}{Representation} \\
\cline{2-7}
\rule{0pt}{3ex}			& $S_4$ & $SO(10)$ &  $ \mathbb{Z}^R_4$&$\mathbb{Z}_{4}$ &$\mathbb{Z}_{4}$ & $\mathbb{Z}_4$ \\ [0.75ex]
\hline \hline
\rule{0pt}{3ex}%
\label{tab:aligfields}
$X_{3^\prime}$ & $3^\prime$ &1 &2 & $0$ &$0$&2 \\
$X_{2}$ & 2 &1 &2 & $2$ &$0$&2 \\
$\tilde{X}_{2}$ & 2 &1 &2 & $0$ &$1$&1 \\
$X_{1}$ & 1 &1 &2 & $0$ &$2$&2 \\
$\tilde{X}_{1}$ & 1 &1 &2 & $3$ &$3$&0 \\
$X_{1'}$ & $1^\prime$ &1 &2 & $3$ &$2$&2 \\
$Z_{3^\prime}$ & $3^\prime$ &1 &2 & $3$ &$0$&2 \\
$\tilde{Z}_{3^\prime}$ & $3^\prime$ &1 &2 & $2$ &$2$&0 \\
$\tilde{Z}$ & 1 &1 &2 & $3$ &$2$&3 \\
$Z$ & 1 &1 &2 & $0$&$0$&0 \\
\hline
\end{tabular}
\caption{Alignment superfields.}
\end{subtable}
\caption{Fields that appear only at high energies. Together with the ones in Table \ref{tab:funfields} they list the complete field content of the model.}
\label{tab:mesfields}
\end{table}

\begin{figure}[ht]
\centering
	\includegraphics[scale=0.7]{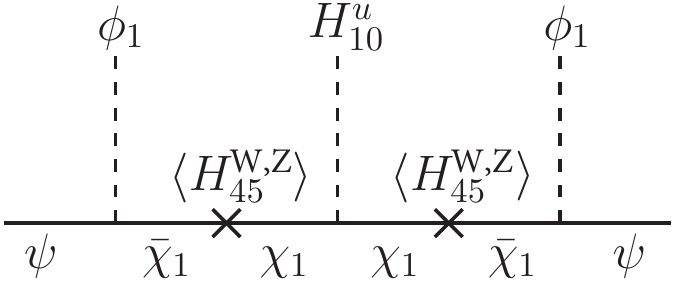}
\hspace*{1ex}
	\includegraphics[scale=0.7]{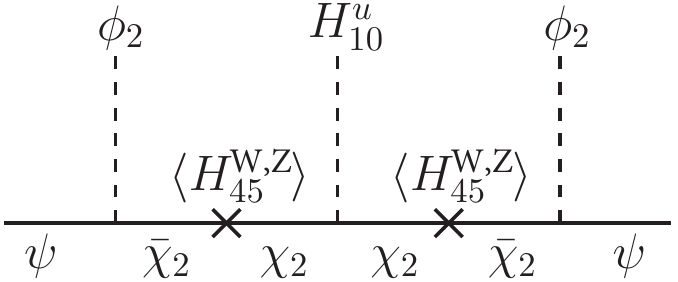}
\hspace*{1ex}
	\includegraphics[scale=0.7]{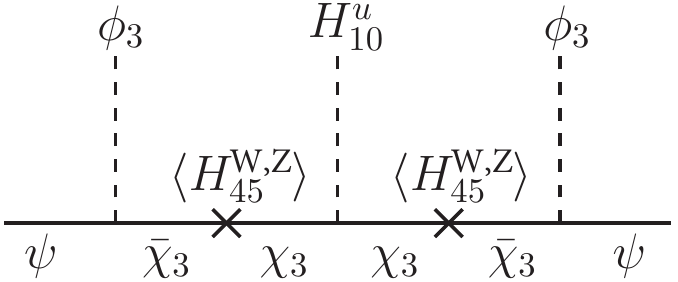}
\caption{Diagrams coupling $ \psi $ to $ H^u_{10} $. When flavons acquire VEVs, these give the up-type quark and Dirac neutrino Yukawa matrices.}
\label{fig:hu}
\end{figure}

\begin{figure}[ht]
\centering
	\includegraphics[scale=0.7]{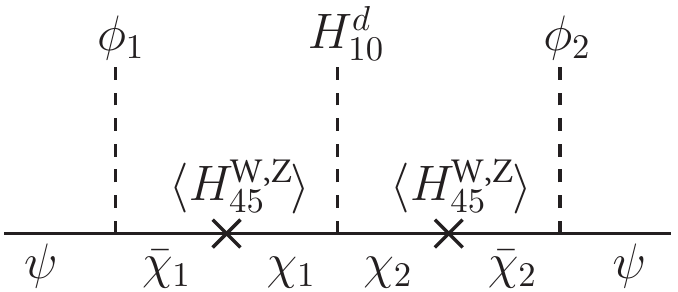}
\hspace*{1ex}
	\includegraphics[scale=0.7]{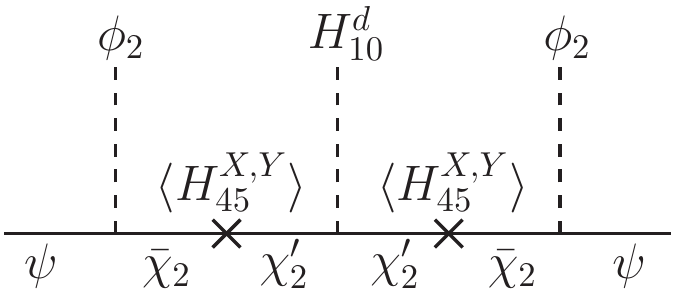}
\hspace*{1ex}
	\includegraphics[scale=0.7]{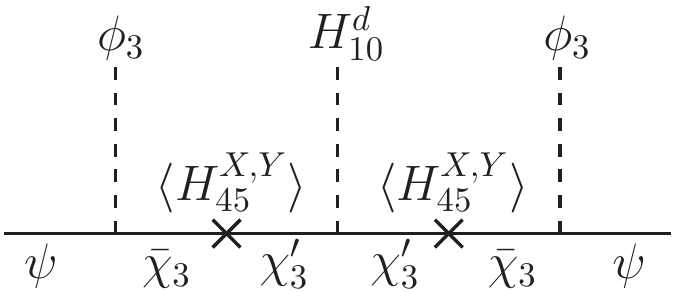}
\caption{Diagrams coupling $ \psi $ to $ H^d_{10} $. These generate the down-type quark and charged lepton Yukawa matrices.}
\label{fig:hd}
\end{figure}

\begin{figure}[ht]
\centering
	\includegraphics[scale=0.7]{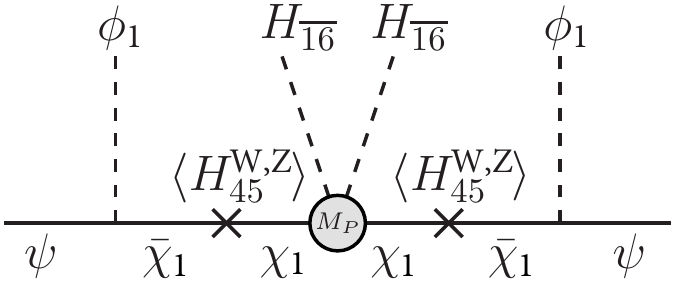}
\hspace*{1ex}
	\includegraphics[scale=0.7]{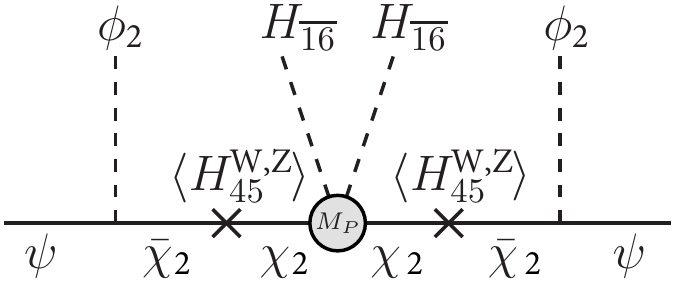}
\hspace*{1ex}
	\includegraphics[scale=0.7]{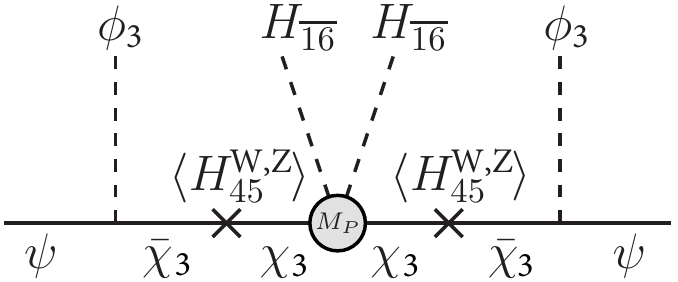}
\caption{Diagrams coupling $ \psi $ to $ H_{\overbar{16}} $. These give the RH neutrino mass matrix.}
\label{fig:h16}
\end{figure}

We now present the effective Yukawa terms in more detail
than in the previous section.  With the fields in the table \ref{tab:funfields} we may write the superpotential relevant to the Yukawa terms, including terms $O(1/M_P)$, as
\begin{equation}
\begin{split}
W_Y
&\sim \frac{H_{10}^u(\psi {\phi_1})(\psi{\phi_1})}{ \braket{H_{45}^{W,Z}}^2}
+ \frac{H_{10}^u(\psi {\phi_2})(\psi{\phi_2})}{ \braket{H_{45}^{W,Z}}^2}
+ \frac{H_{10}^u(\psi {\phi_3})(\psi{\phi_3})}{ \braket{H_{45}^{W,Z}}^2}
\\ 
&+ \frac{H_{10}^d(\psi {\phi_1})(\psi{\phi_2})}{ \braket{H_{45}^{W,Z}}^2}
+ \frac{H_{10}^d(\psi {\phi_2})(\psi{\phi_2})}{ \braket{H_{45}^{X,Y}}^2}
+ \frac{H_{10}^d(\psi {\phi_3})(\psi{\phi_3})}{ \braket{H_{45}^{X,Y}}^2}
\\
&+ \frac{H_{\overbar{16}}H_{\overbar{16}} (\psi {\phi_1})(\psi{\phi_1})}{ M_P\braket{H_{45}^{W,Z}}^2}
+ \frac{H_{\overbar{16}}H_{\overbar{16}} (\psi {\phi_2})(\psi{\phi_2})}{ M_P\braket{H_{45}^{W,Z}}^2}
+ \frac{H_{\overbar{16}}H_{\overbar{16}} (\psi {\phi_3})(\psi{\phi_3})}{ M_P\braket{H_{45}^{W,Z}}^2}
\\
& + \frac{H_{10}^d(\psi\psi)_{3'}(\phi_3)}{M_P}
\label{eq:efyu}
\end{split}
\end{equation}
where $ (\ )_{3'}$ means a $3'$ contraction, while  $(\ )$ without any subscript means the singlet contraction of $S_4$.  There are plenty of terms supressed by $M_P^2$ and they are expected to make small mass contributions of $O(M_{GUT}^2/M_P^2)<10^{-6}$, and therefore negligible \footnote{The most important correction, of $O(10^{-6})$, is made to the up-quark Yukawa coupling. From table \ref{tab:parameters}, we see that it is of comparable magnitude. We performed the fit ignoring these corrections. If they were included, they would shift the fit parameters. The largest contribution to the electron Yukawa coupling is of $O(10^{-8})$ and therefore negligible.}. We have ignored all the $O(1)$ dimensionless couplings for simplicity. The diagrams that generate these terms are shown in Figs.~\ref{fig:hu}-\ref{fig:h16}, where they include the messengers $\chi$, listed in Table \ref{tab:mesfields}. In the Sec. \ref{sec:yukd} we present them in full detail together with the specific messenger structure.

The full field content of the model is listed in Tables~\ref{tab:funfields}-\ref{tab:mesfields}. Even though the list of fields seems large, it is substantially smaller than previous flavoured GUT models that attempt to be complete \cite{Bjorkeroth:2015uou}. 

\section{Vacuum alignment}
\label{sec:vaal}
The flavon superpotential fixes the symmetry breaking flavon VEVs in Eq.~\ref{eq:phfla}. To derive this alignment we use a set of driving fields, listed in Table~\ref{tab:mesfields}, coupled to the flavon fields in Table~\ref{tab:funfields}. We follow a sequence of steps using supersymmetric F-terms equations to align all the flavons. The letter subscript in the flavons refers to the symmetry preserving generator. The alignments depend on the $S_4$ representation of the alignment field, denoted by its index. The superpotential is given by
\begin{equation}\begin{split}
W_\phi&\sim X_{3^\prime}(\phi_{S,U})^2+X_2(\phi_T)^2+X_1(\phi_t)^2+\tilde{X}_{1}\phi_T\phi_t+X_{1'}\phi_T\phi_3+\tilde{X}_2\phi_t\phi_3\\
&\quad +Z_{3^\prime}(\phi_{S,U}\phi_T+\xi \phi_2)+\tilde{Z}_{3^\prime}\xi\left(\frac{\phi_2\phi_3}{M_P}- \phi_1\right),
\end{split}
\label{eq:align potential}
\end{equation}
where we have ignored dimensionless $O(1)$ parameters since they are not relevant. Solving the F-term equations from the alignment fields fixes the flavon VEV alignment, while the F-term equations from flavons forbid the alignment fields from getting a VEV. 

The three $S_4$ generators, working in the $T$ diagonal basis, are 
\begin{equation}
S=\frac{1}{3}
\begin{pmatrix}
 -1 & 2 & 2 \\
 2 & -1 & 2 \\
2 & 2 & -1
\end{pmatrix},\ \ \ \ 
T=
\begin{pmatrix}
 1 & 0 & 0 \\
 0 & \omega^2 & 0 \\
0 & 0 & \omega
\end{pmatrix}
\ \ \ \ { \rm for} \ {\bf{3}} \   { \rm or} \  {\bf{3^\prime}}  \ ,
\label{eq:S4gens}
\end{equation}
and
\begin{equation}
U=\mp
\begin{pmatrix}
 1 & 0 & 0 \\
 0 & 0 & 1 \\
0 & 1 & 0
\end{pmatrix},\ \ \ \ 
SU=US=\mp \frac{1}{3}
\begin{pmatrix}
 -1 & 2 & 2 \\
 2 &  2 & -1 \\
2  & -1 & 2
\end{pmatrix}, \ \ \ \ { \rm for } \ \ {\bf{3}}, {\bf{3^\prime}} \ \  { \rm respectively. }
\end{equation}

The first 3 terms in the superpotential in Eq. \ref{eq:align potential} fix the alignments
\begin{eqnarray}
X_{3^\prime} (\phi_{S,U})^2 & ~~~~~\longrightarrow~~~~~ & 
\begin{pmatrix} 1\\\omega^n\\\omega^{2n}\end{pmatrix}\ ,\\
X_2 (\phi_T)^2 & ~~~~~\longrightarrow ~~~~~ & \begin{pmatrix} 1\\0\\0\end{pmatrix},
\begin{pmatrix} 1\\-2\omega^n \\-2\omega^{2n}\end{pmatrix} \ ,\\
X_1 (\phi_{t})^2 & ~~~~~\longrightarrow~~~~~ & 
\begin{pmatrix} 0\\0\\1\end{pmatrix},
\begin{pmatrix} 0\\1\\0\end{pmatrix},
\begin{pmatrix} 2\\2x\\-1/x\end{pmatrix}\ ,
\end{eqnarray}

 up to an integer ($n \in \mathbb{Z} $) or continuos ($x \in \mathbb{R}$) parameter, with $\omega= e^{2\pi i/3}$. We may notice that the three solutions for $\braket{\phi_{S,U}}$ are related one to another by a $T$ transformation. We may choose it to be $(1,1,1)^T$ without loss of generality.

The $\braket{\phi_T}$ has four different solutions. The last three solutions are related by a $T$ transformation. From these three, the one without any $\omega$ is related to the first solution by an $S$ transformation. Since they are all related, we may choose $(1,0,0)^T$ without loss of generality.

The $\braket{\phi_t}$ has three different solutions.  The third solution is not related to the first two by any symmetry transformations. The fourth term in the superpotential fixes the solution to be either $(0,0,1)^T$ or  $(0,1,0)^T$, which are related by an $U$ transformation and we choose the former without loss of generality.

The fifth and sixth terms fix $\phi_3$ to be orthogonal to $\phi_t$ and $\phi_T$ so that it is fixed to be $(0,1,0)^T$.

The first term from the second line in Eq. \ref{eq:align potential} involves
\begin{equation}
\left(\langle \phi_{S,U} \rangle \cdot \langle \phi_T \rangle \right)_{3^\prime} 
~~ \propto~~
 \begin{pmatrix}0\\-1\\1 \end{pmatrix} ,
\end{equation}
and together with the fifth one fixes $\braket{\phi_2}$ into this direction.
The third term in the second line involves
\begin{equation}
\left(\langle \phi_{2} \rangle\cdot \langle \phi_3 \rangle
  \right)_{3^\prime} ~~ \propto~~ 
 \begin{pmatrix}1\\0\\2 \end{pmatrix} ,
\end{equation}
and together with the seventh term we fix $\braket{\phi_1}$ into this direction. Furthermore the $\xi$ field that does not add anything to the alignment but it plays a role in the VEV driving as explained below.

The F-term equations from the $X,Z$ fix the alignments to be
\begin{equation}\begin{split}
\braket{\phi_{S,U}}=v_1\left(\begin{array}{c}1\\1\\1\end{array}\right),\ \braket{\phi_T}=v_2\left(\begin{array}{c}1\\0\\0\end{array}\right),\ \braket{\phi_t}=v_t\left(\begin{array}{c}0\\0\\1\end{array}\right)\\
\braket{\phi_1}=v_1\left(\begin{array}{c}1\\0\\2\end{array}\right),\ \braket{\phi_2}=v_2\left(\begin{array}{c}0\\1\\-1\end{array}\right),\ \braket{\phi_3}=v_3\left(\begin{array}{c}0\\1\\0\end{array}\right),\ 
\end{split}\end{equation}
where the last three flavons couple to the matter superfield $\psi$ and determine the fermion mass matrix structure. The flavon VEVs $v_i$ are, in general, complex, and spontaneously break the assumed CP symmetry of the high energy theory.

\section{Symmetry breaking}
\label{sec:sb}
The model gives a natural understanding of the SM fermion masses through the hierarchy between the flavon VEVs $|v_1|\ll |v_2|\ll |v_3|$. Here, we show the symmetry breaking superpotential that produces such hierarchy between the VEVs, 
\begin{equation}\begin{split}
\mathcal{W}_{DV}&\sim \tilde{Z}_3\xi\left(\phi_1-\frac{\phi_2\phi_3}{M_P}\right)
\ + \ \tilde{Z}\frac{\phi_{T}}{M_P}\left(\phi_1\phi_2-\frac{\phi_3\sum_i\phi^2_i}{M_P}+O(1/M_P^2)\right)
\\ &+Z\left(M_{GUT}^2+\sum_i\phi_i^2+(H_{45}^{W,Z})^2+(H_{45}^{B-L})^2+\zeta^2+Z^2+O(1/M_P)\right)
\\ &+H_{45}^{B-L}\left((H_{45}^{X,Y})^2+\frac{\zeta}{M_P}\big((H_{45}^{W,Z})^2+(H_{45}^{B-L})^2\big)+H_{45}^{X,Y}\frac{H_{\overbar{16}}H_{16}}{M_P}+DT+O(1/M_P^2)\right),
\label{eq:dr}
\end{split}
\end{equation}
where we have ignored dimensionless couplings for simplicity.

The first term of Eq. \ref{eq:dr} also appears in the alignment potential in Eq.~\ref{eq:align potential} and fixes
\begin{equation}
|\tilde\kappa_1 v_1|= \left|\frac{v_2 v_3}{M_P}\right|,
\label{eq:fixVEV1}
\end{equation}
where $\tilde\kappa_1$ denotes an effective dimensionless coupling coming from the ones in the superpotential. Note that we have written this equation as only fixing the modulus. This happens due to the appearance of the field $\xi$; We assume there are two copies of that field, which get a VEV with an arbitrary phase. This phase, together with the dimensionless couplings for each term, does not allow to relate the phases of the $v_i$.
 
The second term of Eq. \ref{eq:dr} fixes the VEVs
\begin{equation}
\tilde\kappa_2\ v_1 v_2 = \frac{v_3}{M_P}\sum_i v_i^2,
\label{eq:fixVEV2}
\end{equation}
where $\tilde\kappa_2$ denotes an effective dimensionless coupling coming from the ones in the superpotential.
This equation, together with the previous one, require a hierarchy in the $v_i$'s. Specifically it requires $v_{2,3}\gg v_{1}$.

The field $\tilde{Z}$ does not obtain a VEV to comply with the F-term equations from the flavons.

The second line of Eq. \ref{eq:dr} drives the linear combination
\begin{equation}
M_{GUT}^2\sim\sum_i v_i^2+\braket{H_{45}^{W,Z}}^2+\braket{H_{45}^{B-L}}^2+\braket{\zeta}^2+\braket{Z}^2,
\label{eq:drmg}
\end{equation}
where we assume that the sum of $v_i$ and the all adjoints get a GUT scale VEV. The field $Z$ does not get a VEV due to the F-term equations coming from the adjoints. This equation does not fix the phases of the VEVs. We assume that the $\braket{H_{45}^{W,Z}}$ are real while the phase of the sum of flavon VEVs is unconstrained (only related to the one of $\braket{\zeta}$ which does not appear at low energies). We assume that the flavons obtain a VEV that break the CP symmetry with an arbitrary phase.

The third line of Eq. \ref{eq:dr} drives
\begin{equation}
\frac{\braket{\zeta}}{M_P}\big(\braket{H_{45}^{W,Z}}^2+\braket{H_{45}^{B-L}}^2\big)\sim\braket{H_{45}^{X,Y}}^2+\braket{H_{45}^{X,Y}}\frac{\braket{H_{\overbar{16}}H_{16}}}{M_P},
\end{equation}
where we assume that the $\braket{H_{45}^{X,Y}}$ is real. The DT represent all the terms involved in the D-T splitting (shown in Sec.\ref{sec:dt}) that do not contribute to the F-term equation, but they are there nonetheless. The F-term equations coming from the $H_{45}^{X,Y}$ force the $\chi_{u,d}$ to also get a VEV and does not change any low energy phenomenology.

The F-term equations previously discussed can give a VEV to the adjoint fields but do not fix their direction. The adjoint fields can get a VEV in any SM preserving direction. In general they can be written as a linear combination of the $U(1)_{X,Y}$ directions. We do not assume any specific direction for the VEVs $\braket{H_{45}^{W,X,Y,Z},\zeta}$. We assume that $\braket{H_{45}^{B-L}}$ lies in the $U(1)_{B-L}$ direction. \footnote{It can be written as the linear combination $B-L=(-X+4Y)/5$.} We assume that the $\braket{H_{\overbar{16},16}}$ lie in the right handed neutrino direction.

Using the first three equations ~\ref{eq:fixVEV1}-~\ref{eq:drmg}, we may find that the flavon VEVs
\begin{equation}
v_1= \frac{\tilde\kappa_3^2 M_{GUT}^2}{\sqrt{\tilde\kappa_1\tilde\kappa_2} M_{P}}\ v_2,\ \ \ v_2=\frac{\sqrt{\tilde\kappa_1}\tilde\kappa_3M_{GUT}}{\sqrt{\tilde\kappa_2}},\ \ \ v_3= \tilde\kappa_3M_{GUT},
\end{equation}
in this way, if we assume that $\tilde\kappa_1\simeq 0.1,\ \tilde\kappa_2\simeq 10,\ \tilde\kappa_3\simeq 1 $, we have
\begin{equation}
v_3\simeq M_{GUT},\ \ v_2\simeq 0.1\ M_{GUT}, \ \ v_1\simeq 0.001\ M_{GUT},
\end{equation}
which generates the hierarchy between the fermion families. We note that the hierachy between $v_1$ and $v_2$ is given by the structure of the F-term equations. The hierarchy between $v_2$ and $v_3$ is assumed and realized by a much milder hierarchy between the couplings in the superpotential.

Using Eq.~\ref{eq:fixVEV2} and knowing that $v_3\gg v_{1,2}$, we approximately get
\begin{equation}
\tilde\kappa_2\ v_1 v_2 \simeq \frac{v_3^3}{M_P},
\end{equation}
which also fixes the VEV phases to be
\begin{equation}\begin{split}
\arg v_1 + \arg v_2 &\simeq 3 \arg v_3,
\label{eq:argvi}
\end{split}\end{equation}
that in terms of the physical phases is
\begin{equation}
\eta\simeq 4\eta'-2\gamma,
\end{equation}
this way there are only 2 free physical phases.

\section{Doublet-Triplet splitting}
\label{sec:dt}

The Higgs fields $H^{u,d}_{10}$ and $H_{16,\overbar{16}}$ contain $SU(2)$ doublets and $SU(3)$ triplets. We need the triplets to be heavy since they mediate proton decay, while two of the doublets remain light so they can be associated to the MSSM Higgs doublets. This is known as the doublet-triplet splitting problem and can be solved using the Dimopolous-Wilczek mechanism~\cite{DW}. In our case this mechanism is in place since we assume that $\braket{H^{B-L}_{45}}$ lies in the $U(1)_{B-L}$ direction. Furthermore, there are extra pairs of doublets, and they are required to be heavy to preserve gauge coupling unification. Using the fields in Tables \ref{tab:funfields}-\ref{tab:mesfields}, we may write the superpotential involving the Higgs fields (ignoring dimensionless parameters)
\begin{equation}\begin{split}
\mathcal{W}_H &= 
	H^{B-L}_{45} \left(H^u_{10}H^d_{10}+\zeta_2 \zeta_2+H_{\overbar{16}}\chi_u+H_{16}\overbar{\chi}_d\right)\\
	&\quad +H_{\overbar{16}}H_{10}^u\overbar{\chi}_u+H_{16}H_{10}^d\chi_d+H_{16}H_{\overbar{16}}\zeta_1	
	% \\ &\quad
	+\zeta\left(\zeta_1\zeta_2+\overbar{\chi}_u\chi_u+\overbar{\chi}_d\chi_d\right) \\ 
	&\quad 
	+ H^{B-L}_{45}
	\left( 
	\frac{H_{\overbar{16}}H_{\overbar{16}}H^d_{10}}{M_P}
	+\frac{H_{16}H_{16}H^u_{10}}{M_P}
	+H_{10}^uH_{10}^d\frac{(H_{45}^{X,Y,W,Z})^4}{M_P^4}
	\right).
\end{split}\end{equation}
After integrating out the messengers $\zeta_i,\chi_j$, the superpotential becomes 
\begin{equation} \begin{split}
\mathcal{W}_H &=H^{B-L}_{45}\left(\kappa_1H^u_{10}H^d_{10}+\kappa_2\frac{(H_{16}H_{\overbar{16}})^2}{\braket{\zeta}^2}+\kappa_7H_{10}^uH_{10}^d\frac{(H_{45}^{X,Y,W,Z})^4}{M_P^4}\right.\\
&\hspace{2cm}+\kappa_3\frac{H_{\overbar{16}}H_{\overbar{16}}H^u_{10}}{\braket{\zeta}}+\kappa_4\frac{H_{\overbar{16}}H_{\overbar{16}}H^d_{10}}{M_P}
+\left. \kappa_5\frac{H_{16}H_{16}H^u_{10}}{M_P}+\kappa_6\frac{H_{16}H_{16}H^d_{10}}{\braket{\zeta}}\right).
\end{split} \end{equation}
We remember that the magnitude of the VEVs is
\begin{equation}
\braket{H_{16}}\simeq \braket{H_{\overbar{16}}} \simeq \braket{H_{45}}= M_{GUT},
\end{equation}
and we define
\begin{equation}
z= M_{GUT}/\braket{\zeta}, \ \ \ y=M_{GUT}/M_P.
\end{equation}

Denoting the up (down)-type doublet inside each $H_{10}$ as $\textbf{2}_{u(d)}(H^{u,(d)}_{10})$, and similarly for the triplets, the mass matrix for the triplets becomes
\begin{equation}
	M_T\sim \begin{blockarray}{c ccc}
	& \textbf{3}_{u}(H^u_{10}) &  \textbf{3}_{u}(H^d_{10}) &  \textbf{3}_{u}(H_{\overbar{16}})\\[1ex]
	\begin{block}{c(ccc)}
	 \textbf{3}_{d}(H^d_{10}) & \kappa_1 &0& \kappa_4 y \\[0.5ex]
	 \textbf{3}_{d}(H^u_{10}) &0 & -\kappa_1 & \kappa_3 z \\[0.5ex]
	 \textbf{3}_{d}(H_{16}) &\kappa_5 y  & \kappa_6 z & \kappa_2  z^2\\
	\end{block}
	\end{blockarray}\ M_{GUT},
\label{eq:trmass}
\end{equation}
that has as approximate eigenvalues
\begin{equation}
m_T\sim\ \ \kappa_1  M_{GUT},\ \ \kappa_1  M_{GUT},\ \ \kappa_2 z^2 M_{GUT},
\end{equation}
so that it requires
$
\kappa_1\sim \kappa_2 z^2 \sim 1,
$
to get the triplets at the GUT scale.
The doublets mass matrix is
\begin{equation}
	M_D\sim \begin{blockarray}{c ccc}
	& \textbf{2}_{u}(H^u_{10}) &  \textbf{2}_{u}(H^d_{10}) &  \textbf{2}_{u}(H_{\overbar{16}})\\[1ex]
	\begin{block}{c(ccc)}
	 \textbf{2}_{d}(H^d_{10}) & -\kappa_7 y^4&0& \kappa_4 y \\[0.5ex]
	 \textbf{2}_{d}(H^u_{10}) &0 & \kappa_7 y^4  & \kappa_3 z \\[0.5ex]
	 \textbf{2}_{d}(H_{16}) &\kappa_5 y  & \kappa_6 z & \kappa_2 z^2 \\
	\end{block}
	\end{blockarray}\ M_{GUT},\label{eq:dmass}
\end{equation}
that has as eigenvalues
\begin{equation}
m_D\sim \ \ -y^4 M_{GUT}, \ \  \kappa_6\kappa_3 z^2 M_{GUT},\ \ \kappa_2 z^2 M_{GUT},
\end{equation}
so that we must have
$
\kappa_6\kappa_3 z^2\sim \kappa_2 z^2  \sim 1,
$
to get two doublet pairs at the GUT scale. Furthermore, there is a $\mu$ term generated by
\begin{equation}
\mu\sim y^4 M_{GUT} \sim 1\ TeV,
\end{equation}
which happens at the correct order.

The light MSSM doublets are
\begin{equation}
\begin{split}
h_u \simeq \textbf{2}_{u}(H^u_{10})+\frac{\kappa_4 y}{\kappa_3 z}\textbf{2}_{u}(H^d_{10}),\hspace{7mm}
h_d \simeq \textbf{2}_{d}(H^d_{10})+\frac{\kappa_5 y}{\kappa_6 z}\textbf{2}_{u}(H^d_{10}),
\end{split}
\end{equation}
so that the second term is suppressed to be $<10^{-3}$ and we may safely assume that $h_{u(d)}$ lies only inside $H^{u(d)}_{10}$.

\section{Proton decay}
\label{sec:pd}

One of the characteristic features of GUTs is the prediction of proton decay. 
It has not been observed and the proton lifetime is constrained to be $\tau_p > 10^{34} $ years \cite{Olive:2016xmw}. 

Proton decay can be mediated by the extra gauge bosons of the GUT and by the triplets accompanying the Higgs doublets. 
In SUSY $SO(10)$ GUTs, the main source for proton decay comes from the triplet Higgsinos. 
The decay width is dependent on SUSY breaking and the specific coupling texture of the triplets and determining it exactly lies beyond the scope of this paper. 
In general the constraints are barely met when the triplets have a mass at the GUT scale \cite{Murayama:1994tc}, and in Sec. \ref{sec:dt} we have shown this is our case. 

The existence of additional fields in the model may allow proton decay to arise from effective terms of the type
\begin{equation}
	gQQQL\frac{\braket{X}}{M_P^2}.
\end{equation}
Such terms must obey the constraint $ g \braket{X} < 3\times 10^{9} $ GeV \cite{Murayama:1994tc}.
In our model, the largest contribution of this type comes from the term
\begin{equation}
	\psi\psi\psi\psi
	\frac{\braket{H^{B-L}_{45}(H_{45}^{X,Y})^2}}{M^4_P}
	\quad \Rightarrow \quad 
	\braket{X} = \frac{(M_\mathrm{GUT})^3}{M^2_P}\sim 10^{10} \mathrm{~GeV}.
\end{equation}
The constraints are met when $g<0.3$. With an $O(1)$ $g$ parameter, the contributions coming from these terms are the same order as the ones coming from the Higgs triplets. In this model, proton decay complies with experimental constraints but lies fairly close to detection.

\section{Detailed Yukawa structure}
\label{sec:yukd}

Now that we have given VEVs to the fields in a specific direction, we may write the fully detailed Yukawa structure.

With the fields in the Table \ref{tab:funfields}, together with the messenger fields in Table \ref{tab:mesfields} we may write the superpotential relevant to the Yukawa terms, up to $O(1/M_P)$,
\begin{equation}
\begin{split}
W_Y&= \sum_{a=1,2,3}\left( \lambda^\phi_a\left(\psi\phi_a\right)\bar{\chi}_a+(\lambda^W_a H_{45}^W+\lambda^Z_a H_{45}^Z)\chi_a\bar{\chi}_a +\lambda^u_a \chi_a\chi_aH^u_{10}+
\lambda^N_a\chi_a\chi_a\frac{H_{\overbar{16}}H_{\overbar{16}}}{M_P}\right)
\\ &\quad +\sum_{b=2,3}\Big( \overbar{\chi}_b\chi_b^d(\lambda^{X}_aH_{45}^X+\lambda^{Y}_aH_{45}^Y)+\lambda^d_b\chi^d_b\chi^d_bH_{10}^d \Big)+\lambda_{12}^d\chi_1\chi_2H^d_{10}+\lambda^d_t\frac{\left(\psi\psi \right)_{3^\prime}\phi_3H^d_{10}}{M_P},
\label{eq:wy1}
\end{split}
\end{equation}
where $\left(\ \right),\ \left(\ \right)_{3^\prime}$ means an $S_4$ singlet or $3^\prime$ contraction respectively. The $\lambda$'s are dimensionless and real coupling constants (due to CP conservation) and are all expected to be $O(1)$.  

After integrating the messengers $\chi$, we obtain the superpotential
\begin{equation}
\begin{split}
W_Y&=\sum_{a=1,2,3}\left( \frac{(\lambda^\phi_a)^2\left(\psi\braket{\phi_a}\right) \left(\psi\braket{\phi_a}\right)}{(\lambda^W_a \braket{H_{45}^W}+\lambda^Z_a \braket{H_{45}^Z})^2}\lambda^u_\chi H_{10}^u+\frac{(\lambda^\phi_a)^2\left(\psi\braket{\phi_a}\right)\left(\psi\braket{\phi_a}\right)}{(\lambda^W_a \braket{H_{45}^W}+\lambda^Z_a \braket{H_{45}^Z})^2}\frac{\lambda^N_a}{M_P}\braket{H_{\overbar{16}}}\braket{H_{\overbar{16}}}\right)
\\ &+\left(\sum_{b=2,3}\lambda^u_a\frac{(\lambda^\phi_b)^2\left(\psi\braket{\phi_b}\right)\left(\psi\braket{\phi_b}\right)}{(\lambda_b^X\braket{H_{45}^X}+\lambda_b^Y\braket{H_{45}^Y})^2}+\lambda_{12}^d\frac{\lambda^\phi_1\lambda^\phi_2\left(\psi\braket{\phi_1}\right)\left(\psi\braket{\phi_2}\right)}{(\lambda^W_1 \braket{H_{45}^W}+\lambda^Z_1 \braket{H_{45}^Z})(\lambda^W_2 \braket{H_{45}^W}+\lambda^Z_2 \braket{H_{45}^Z})}\right.\\
&\left.+\quad\lambda^d_t\frac{\left(\psi\psi\right)_{3^\prime}\braket{\phi_3}}{M_P} \right) H_{10}^d.
\label{eq:efyu}
\end{split}
\end{equation}
This superpotential generates all the SM fermion masses.

\subsection{Renormalisability of the third family}

In Eq. \ref{eq:efyu}, all the terms suppressed by $\braket{H_{45}^{X,Y,W,Z}}$ involve integrating out the messengers by assuming $M_{GUT}\gg v_i$. This naive integration is not possible for the third flavon since it has a much larger VEV $v_3\sim M_{GUT}$. 
Let us single out the terms in $ W_Y $ involving these fields. 
Ignoring $ \mathcal{O}(1) $ couplings, and after the fields get their VEV, the relevant terms are
\begin{equation}
	W_Y^{(3)} \sim 
	 v_3 \psi_3 \overbar{\chi}_3 + \braket{H_{45}^{W,Z}} \chi_3 \overbar{\chi}_3 .
\label{eq:WY3}
\end{equation}
Naively,  one would interpret $ \psi_3 $ as the set of third-family particles, but the first term in Eq.~\ref{eq:WY3} generates mixing with $\overbar{\chi}_3$. 
To obtain the physical (massless) states, which we label $ t $, we rotate into a physical basis $ (\psi_3,\chi_3) \to (t,\chi) $
\begin{equation}
\begin{split}
	\psi_3 = 
	\frac{\braket{H_{45}^{W,Z}} t + v_3 \, \chi}{r}
	,
	\quad
	\chi_3 = 
	\frac{- v_3 \, t + \braket{H_{45}^{W,Z}} \chi}{r}
	;
	\quad
	r = \sqrt{v_3^2+\braket{H_{45}^{W,Z}}^2 }.
\end{split}
\end{equation} 
Physically, it may be interpreted as follows: inside the original superpotential $ W_Y $ lie the terms
\begin{equation}
	\mathcal{W}_Y 
	\supset \chi_3 \chi_3 H^{u,d}_{10}
	\supset \frac{v_3^2}{v_3^2 + \braket{H_{45}^{W,Z}}^2} \,
	t \, t \, H^{u,d}_{10} , 
\end{equation}
which generate renormalisable mass terms for the third family at the electroweak scale.

\subsection{Mass matrix structure}

The superpotential in Eq.~\ref{eq:efyu} generates all the SM fermion mass matrices. The structure of the mass matrices is fixed by the flavon VEV structure shown in Eq.~\ref{eq:phfla}. We may redefine the dimensionless couplings to obtain the mass structure of the SM fermions at low energies
\begin{equation}
\begin{split}
y^{u}_{a=1,2} &=
		 \lambda^u_a 
		\frac{(\lambda^\phi_a)^2 |v_a|^2}%
		{[\lambda_a^W\braket{H_{45}^W} + \lambda_a^Z \braket{H_{45}^Z}]_Q  [\lambda_a^W \braket{H_{45}^W} + \lambda_a^Z \braket{H_{45}^Z}]_{u^c}},\\
y_3^u&=\frac{(\lambda^\phi_3)^2 |v_3|^2}%
		{(\lambda^\phi_3)^2 v_3^2+[\lambda_3^W\braket{H_{45}^W} + \lambda_3^Z \braket{H_{45}^Z}]_Q  [\lambda_3^W \braket{H_{45}^W} + \lambda_3^Z \braket{H_{45}^Z}]_{u^c}},\\
y^{\nu}_{a=1,2} &=
		 \lambda^u_a 
		\frac{(\lambda^\phi_a)^2 |v_a|^2}%
		{[\lambda_a^W\braket{H_{45}^W} + \lambda_a^Z \braket{H_{45}^Z}]_L  [\lambda_a^W \braket{H_{45}^W} + \lambda_a^Z \braket{H_{45}^Z}]_{\nu^c}},\\
 y_3^\nu&=\frac{(\lambda^\phi_3)^2 |v_3|^2}%
		{(\lambda^\phi_3)^2 v_3^2+(\lambda^\chi_3)^2 [\lambda_3^W\braket{H_{45}^W} + \lambda_3^Z \braket{H_{45}^Z}]_L  [\lambda_3^W \braket{H_{45}^W} + \lambda_3^Z \braket{H_{45}^Z}]_{\nu^c}},\\
	y^{e}_{2} &=
		 \lambda^d_2 
		\frac{(\lambda^\phi_2)^2|v_2|^2}%
		{[\lambda_2^X \braket{H_{45}^X} + \lambda_2^Y \braket{H_{45}^Y}]_L 
		 [\lambda_2^X \braket{H_{45}^X} + \lambda_2^Y \braket{H_{45}^Y}]_{e^c}}, \\
	y^e_3&=
		 \lambda^d_3 
		\frac{(\lambda^\phi_3)^2|v_3|^2}%
		{(\lambda^\phi_3)^2v_3^2+[\lambda_3^X \braket{H_{45}^X} + \lambda_3^Y \braket{H_{45}^Y}]_L
		 [\lambda_3^X \braket{H_{45}^X} + \lambda_3^Y \braket{H_{45}^Y}]_{e^c}}, \\
		y^{d}_{2} &=
		 \lambda^d_2 
		\frac{(\lambda^\phi_2)^2|v_2|^2}%
		{[\lambda_2^X \braket{H_{45}^X} + \lambda_2^Y \braket{H_{45}^Y}]_Q 
		 [\lambda_2^X \braket{H_{45}^X} + \lambda_2^Y \braket{H_{45}^Y}]_{d^c}},\\
	y^d_3&= \lambda^d_3 
		\frac{(\lambda^\phi_3)^2|v_3|^2}%
		{(\lambda^\phi_3)^2v_3^2+[\lambda_3^X \braket{H_{45}^X} + \lambda_3^Y \braket{H_{45}^Y}]_Q 
		 [\lambda_3^X \braket{H_{45}^X} + \lambda_3^Y \braket{H_{45}^Y}]_{d^c}}, \\
	y^{e}_{12}&=
		 \lambda^d_{12} 
		\frac{\lambda^\phi_1\lambda^\phi_2 |v_1 v_2|}%
		{[\lambda_1^W\braket{H_{45}^W} + \lambda_1^Z \braket{H_{45}^Z}]_{L+e^c}  [\lambda_2^W \braket{H_{45}^W} + \lambda_2^Z \braket{H_{45}^Z}]_{L+e^c}},\\
		y^d_{12}&=\lambda^d_{12} 
		\frac{\lambda^\phi_1\lambda^\phi_2 |v_1 v_2|}
		{[\lambda_1^W\braket{H_{45}^W} + \lambda_1^Z \braket{H_{45}^Z}]_{Q+d^c}  [\lambda_2^W \braket{H_{45}^W} + \lambda_2^Z \braket{H_{45}^Z}]_{Q+d^c}},\\
	M^{\mathrm{R}}_{a=1,2} &=  
		 \frac{\lambda^N_a v_{\overbar{16}}^2}{M_P}
		\frac{(\lambda^\phi_a)^2 |v_a|^2}%
		{[\lambda_a^W \braket{H_{45}^W} + \lambda_a^Z \braket{H_{45}^Z}]_{\nu^c}^2},\\
	M^{\mathrm{R}}_{3} &=  \frac{\lambda^N_3 v_{\overbar{16}}^2}{M_P}
		\frac{(\lambda^\phi_3)^2 |v_3|^2}%
		{(\lambda^\phi_3)^2 v_3^2+[\lambda_3^W \braket{H_{45}^W} + \lambda_3^Z \braket{H_{45}^Z}]_{\nu^c}^2},\\
		y^P &= \lambda^d_t \frac{ Y_P v_3}{M_P} , 
\label{eq:yudeta}
\end{split}
\end{equation}
where $\braket{H_{45}^{X,Y,W,Z}}_f$ denotes the adjoint VEV with the corresponding CG coefficients for each SM fermion $f$. This allows for each $y,M$ parameter in Eq. \ref{eq:yudeta} to be independent. The VEVs $\braket{H^{X,Y}}$ obtain a VEV in an arbitrary $SO(10)$ breaking direction. They need to be different from one another. 

For a better understanding we can show an explicit example. Let us assume that  $\braket{H_{45}^{X,Y}}$ is aligned in the $U(1)_{X,Y}$ direction respectively with an $M_{GUT}$ magnitude. In this case the effective Yukawa couplings $y_2^{e,d}$ would be
\begin{equation}
y^{e}_{2} =
		 \lambda^d_2 
		\frac{(\lambda^\phi_2)^2|v_2|^2}%
		{[3\lambda_2^X  - \lambda_2^Y /2]
		 [-\lambda_2^X  + \lambda_2^Y ]M_{GUT}^2},\ \ \ 
y^{d}_{2} =
		 \lambda^d_2 
		\frac{(\lambda^\phi_2)^2|v_2|^2}%
		{[-\lambda_2^X + \lambda_2^Y/6] 
		 [3\lambda_2^X + \lambda_2^Y/3]M_{GUT}^2},
\end{equation}
where the coefficients multiplying each $\lambda^{X,Y}$ are the $U(1)_{X,Y}$ charges of the corresponding SM field. Since the $\lambda_2^{X,Y}$ appear with different coefficients in $y^{e,d}_2$, we can use them to obtain a arbitrarily different effective Yukawa coupling for charged leptons and down type quarks.

Assuming the adjoints have all real VEVs, the physical phases are
\begin{equation}\begin{split} 
\eta&=2\arg v_1-2\arg v_2\\
\eta'&=2\arg v_3-2\arg v_2\\
\gamma&=\arg v_3-2\arg v_2,
\end{split}\end{equation}
while all the $y's$ and $M's$ are real.

With these definitions we may write the fermion mass matrices
	\begin{equation}
\begin{alignedat}{7}
& M^e/ v_d &&=\, & y^e_{12}e^{i\eta/2}\pmatr{0&1&1\\1&4&2\\1&2&0}\,+ \, & y^e_2\pmatr{ 0&0&0 \\ 0&1&1 \\ 0&1&1 }\,+ \, &y^e_3e^{i \eta'}  \pmatr{ 0&0&0 \\ 0&0&0 \\ 0&0&1 }\, +\, &y^P e^{i\gamma}\pmatr{0&0&1\\0&2&0\\1&0&0} ,  \\
&M^d/v_d&&=\, &y^d_{12}e^{i\eta/2}\pmatr{0&1&1\\1&4&2\\1&2&0}\,+\, &y^d_2\pmatr{ 0&0&0 \\ 0&1&1 \\ 0&1&1 }\,+\, &y^d_3e^{i\eta'}\pmatr{ 0&0&0 \\ 0&0&0 \\ 0&0&1}\,+ \, &y^Pe^{i\gamma}\pmatr{0&0&1\\0&2&0\\1&0&0},
\\ 
&M^u/v_u&& = \, &y^u_1e^{i\eta}\pmatr{ 1&2&0 \\ 2&4&0 \\ 0&0&0}\, + \, &y^u_2\pmatr{ 0&0&0 \\ 0&1&1 \\ 0&1&1 } + 
&y^u_3e^{i \eta'}  \pmatr{ 0&0&0 \\ 0&0&0 \\ 0&0&1 } ,
\\ 
&M_D^\nu/v_u&& = \, &y^\nu_1e^{i\eta}\pmatr{ 1&2&0 \\ 2&4&0 \\ 0&0&0}\, + \, &y^\nu_2\pmatr{ 0&0&0 \\ 0&1&1 \\ 0&1&1 } + &y^\nu_3e^{i \eta'}  \pmatr{ 0&0&0 \\ 0&0&0 \\ 0&0&1 } ,
\\ 
& M^{R} &&= \, &M^R_1e^{i\eta}\pmatr{ 1&2&0 \\ 2&4&0 \\ 0&0&0}\, + \, & M^R_2\pmatr{ 0&0&0 \\ 0&1&1 \\ 0&1&1 }\, + \, &M^R_3e^{i \eta'}  \pmatr{ 0&0&0 \\ 0&0&0 \\ 0&0&1 } .
\label{eq:masmat}
\end{alignedat}
\end{equation}
 We note the remarkable universal structure of the matrices in the up and neutrino sectors, 
 which differ from the down and charged lepton sectors.

The $y$ and $M$ parameters are all free and independent while there is a constraint in the phases
\begin{equation}
\eta\simeq 4\eta'-2\gamma,
\end{equation}
as shown in the Sec.~\ref{sec:sb}.
We have in total 18 free parameters that fix the whole spectrum of fermion masses and mixing angles, as discussed in Sec.~\ref{sec:parameter}.

As shown in the Sec.~\ref{sec:sb}, the flavons and adjoint fields get a VEV 
\begin{equation}\begin{split}
M_{GUT}&\sim v_3\sim 10\ v_2\sim 1000\ v_1,\\
M_{GUT}&\sim v_{45}^{X,Y,W,Z}\sim \ M_\rho\sim  v_{\overbar{16}}.
\end{split}\end{equation}
Assuming all the parameters in the superpotential we have ignored are $O(1)$, and $\tan\beta\sim 20$, the mass matrix parameters are expected to be
\begin{equation}\begin{split}
y_1^u\sim y^\nu_1\sim v_1^2/v_{45}^2\sim 10^{-6}, &\ \ \ 
y_2^u\sim y^\nu_2\sim v_2^2/v_{45}^2\sim 10^{-2},\\
y_3^u\sim y^\nu_3\sim v_3^2/v_{45}^2\sim 1,&\ \ \
y_{12}^d\sim y^e_{12}\sim \cos\beta\ v_1v_2/v_{45}^2\sim 10^{-5},\\
y_2^d\sim y^e_2\sim \cos\beta\ v_2^2/v_{45}^2\sim 10^{-3}, &\ \ \
y_3^d\sim y^e_3\sim \cos\beta\ v_3^2/v_{45}^2\sim 0.1,\\
y^P\sim \cos\beta\ v_3/M_P\sim 10^{-4} , &\ \ \
M_1^R\sim v_{\overbar{16}}^2v_1^2/v_{45}^2M_P\sim 10^7\ GeV, \\
M_2^R\sim v_{\overbar{16}}^2v_2^2/v_{45}^2M_P\sim 10^{11}\ GeV, &\ \ \ 
M_3^R\sim v_{\overbar{16}}^2v_3^2/v_{45}^2M_P \sim 10^{13}\ GeV.
\label{eq:natva}
\end{split}\end{equation} 
These values denote only an approximate order of magnitude for each parameter and are expected to be different due to the appearance of dimensionless couplings. This applies specially to the last 4 parameters that come from unknown Planck suppressed physics and may deviate significantly from our naive expectation.

\subsection{Seesaw mechanism}
\label{sec:seesaw} 
Since we have very heavy RHN Majorana masses, the left handed neutrinos get a very small Majorana mass through type I seesaw
\begin{equation}
m^{\nu}_{L}=M_D^\nu (M^{R})^{-1}(M_D^\nu)^T.
\label{eq:ss}
\end{equation}
As we see in Eq. \ref{eq:masmat}, the Dirac neutrino masses $M_D^\nu$
and RHN Majorana masses $M^{R}$ have the same matrix structure. These are rank one matrices so that we may write them as
\begin{equation}\begin{split}
M_D^\nu/v_u&=y^\nu_1 e^{i\eta}\ \varphi_1\varphi_1^T+y^\nu_2 \ \varphi_2\varphi_2^T+y^\nu_3 e^{i\eta'}\ \varphi_3\varphi_3^T,\\
M^{R}&=M^R_1 e^{i\eta}\ \varphi_1\varphi_1^T+M^R_2 \ \varphi_2\varphi_2^T+M^R_3 e^{i\eta'}\ \varphi_3\varphi_3^T,
\end{split}\end{equation}
with
\begin{equation}
\varphi_1^T=(1,2,0),\ \ \ \varphi^T_2=(0,1,1), \ \ \ \varphi^T_3=(0,0,1).
\end{equation}
We may always find vectors $\tilde\varphi_a$ such that 
\begin{equation}
\tilde\varphi^T_i\varphi_j=\delta_{ij},
\end{equation}
this way we may write the inverse matrix as
\begin{equation}
(M^{R})^{-1}=\frac{e^{-i\eta}}{M^R_1}\tilde\varphi_1\tilde\varphi_1^T+\frac{1}{M^R_2}\tilde\varphi_2\tilde\varphi_2^T+\frac{e^{-i\eta'}}{M^R_3}\tilde\varphi_3\tilde\varphi_3^T.
\end{equation}
Plugging this into the seesaw mechanism we obtain the light effective left-handed Majorana 
neutrino mass matrix $m^{\nu}_{L}$,
\begin{equation}\begin{split}
m^{\nu}_{L}&=\mu_1 e^{i\eta}\ \varphi_1\varphi_1^T+\mu_2 \ \varphi_2\varphi_2^T+\mu_3 e^{i\eta'}\ \varphi_3\varphi_3^T, \ \ \ \textrm{with} \ \ \ \mu_a=\frac{(y^\nu_a v_u)^2}{M^R_a}
\label{eq:ssf}
\end{split}\end{equation}
so that we may conclude that the small left handed neutrino mass matrix has the same universal structure
\begin{equation}
 m^{\nu}_{L} = \, \mu_1e^{i\eta}\pmatr{ 1&2&0 \\ 2&4&0 \\ 0&0&0} \, + \, \mu_2\pmatr{ 0&0&0 \\ 0&1&1 \\ 0&1&1 } \, + \, \mu_3e^{i \eta'}  \pmatr{ 0&0&0 \\ 0&0&0 \\ 0&0&1 } ,
\label{eq:mumatr}
\end{equation}
after the seesaw mechanism.

\section{Numerical fit}
\label{sec:fit}
To test our model we perform a numerical fit using a $\chi^2$ test function. We have a set of input parameters $x=\{ y^u_i, y^d_i, y^e_i, y^P, \mu_i, \eta^\prime, \gamma \}$, from which we obtain a set of observables $P_n(x)$. We minimize the function defined as
\begin{equation}
\chi^2 = \sum_n \left( \frac{P_n(x) - P^\mathrm{obs}_n}{\sigma_n} \right)^2,
\end{equation}
where the 19 observables are given by $P^\mathrm{obs}_n \in \{ \theta^q_{ij}, \delta^q, y_{u,c,t}, y_{d,s,b}, \theta^\ell_{ij}, \delta^l, y_{e,\mu,\tau}, \Delta m_{ij}^2 \}$ with statistical errors $\sigma_n$. This test assumes data is normally (Gaussian) distributed, which is true for most of the observables except for $\theta_{23}^\ell$. The atmospheric mixing angle octant, i.e. $\theta_{23}^\ell < 45^{\circ} $ or $\theta_{23}^\ell > 45^{\circ} $, has not been determined yet. Current data favours $\theta_{23}^\ell=41.6$~\cite{Esteban:2016qun} and we assume such scenario.   
\begin{table}[!ht]
	\centering
	\footnotesize
	\renewcommand{\arraystretch}{1.1}
	\begin{tabular}{ l cc c c }
		\toprule
		\multirow{2}{*}{Observable}& \multicolumn{2}{c}{Data} && \multicolumn{1}{c}{Model} \\
		\cmidrule{2-5}
		& Central value & 1$\sigma$ range  && Best fit \\
		\midrule
		$\theta_{12}^\ell$ $/^\circ$ & 33.57 & 32.81 $\to$ 34.33 && 33.53  \\ 
		$\theta_{13}^\ell$ $/^\circ$ & 8.460 & 8.310 $\to$ 8.610 && 8.452 \\  
		$\theta_{23}^\ell$ $/^\circ$ & 41.75 & 40.40 $\to$ 43.10  && 41.88 \\ 
		$\delta^\ell$ $/^\circ$ & 261.0 & 206.0 $\to$ 316.0 && 200.3  \\
		$y_e$  $/ 10^{-5}$ & 6.023 &  5.987 $\to$ 6.059 && 6.023 \\ 
		$y_\mu$  $/ 10^{-2}$ & 1.272 & 1.264 $\to$ 1.280 && 1.272 \\ 
		$y_\tau$  & 0.222 & 0.219 $\to$ 0.225 && 0.222 \\ 
		$\Delta m_{21}^2 / (10^{-5} \, \mathrm{eV}^2 ) $ & 7.510  & 7.330 $\to$ 7.690 && 7.507 \\
		$\Delta m_{31}^2 / (10^{-3} \, \mathrm{eV}^2) $ & 2.524  & 2.484 $\to$ 2.564 && 2.524 \\
		$m_1$ /meV & & && 10.94  \\ 
		$m_2$ /meV & & && 13.95  \\ 
		$m_3$ /meV & & && 51.42 \\
		$\sum m_i$ /meV & \multicolumn{2}{c}{$<$ 230} && 76.31 \\
		$ \alpha_{21} $ $/^\circ$ & & && 134.3 \\
		$ \alpha_{31} $  $/^\circ$& & && 6.415  \\
		$ m_{\beta \beta}$ /meV &  \multicolumn{2}{c}{$<$ 61-165}   && 11.10\\
		\bottomrule     		
	\end{tabular}
	\caption{Model predictions in the lepton sector for $\tan \beta = 20$, $ M_{\mathrm{SUSY}} = 1 $ TeV and $ \bar{\eta}_b = -0.9 $. The observables are at the GUT scale. The lepton contribution to the total $\chi^2$ is 1.2. The neutrino masses $m_i$ as well as the Majorana phases are pure predictions of our model. The bound on $ \sum m_i $ is taken from\cite{Ade:2015xua}.
	The bound on $ m_{\beta \beta}$  is taken from~\cite{KamLAND-Zen:2016pfg}.
	}
	\label{tab:numleptons}
\end{table}

We need to run up all the measured Yukawa couplings and mixing angles up to the GUT scale in order to compare it with the predictions of our model. 
\footnote{Note that we are performing the numerical fit in terms of the effective neutrino mass parameters $\mu_i$ defined in Eq. \ref{eq:mumatr}. We are ignoring any renormalisation group running corrections in the neutrino sector.}
In doing so, we need to match the SM to the MSSM at the SUSY scale, $M_{SUSY}$, which involves adding the  supersymmetric radiative threshold corrections. This has been done in \cite{Antusch:2013jca}. At the GUT scale, the values depend to a good approximation only on $\bar{\eta}_b$ and $\tan \beta$. A good fit is found for large $\bar{\eta}_b $, which can be explained if $\tan \beta \gtrsim 10$, as shown in the Sec.~\ref{sec:susy}. We also need $\tan \beta < 30$ to keep Yukawa couplings perturbative. The best fit is found for $\bar{\eta}_b=-0.9$ and $\tan \beta=20$. The SUSY scale does not affect the fit and we choose $M_{SUSY}=1$ TeV. The fit has been performed using the Mixing Parameter Tools (MPT)  package~\cite{Antusch:2005gp}.

The best fit found has a $\chi^2=11.9$. Table \ref{tab:numleptons} shows the best fit to the charged leptons and neutrinos observables. Neutrino data is taken from the Nufit global fit \cite{Esteban:2016qun}. Only the neutrino mass-squared differences are known but our model also predicts the neutrino masses themselves as well as the Majorana phases. The model predicts normal ordered neutrino masses and we also give the effective Majorana mass $m_{\beta\beta}$. All the lepton sector is fitted to within $1\sigma$ except for the leptonic CP phase. $\delta^\ell$ is not yet well measured, although a negative CP phase is preferred \cite{Abe:2017uxa}.
\begin{table}[!ht]
	\centering
	\footnotesize
	\renewcommand{\arraystretch}{1.1}
	\begin{tabular}{ l cc c c }
		\toprule
		\multirow{2}{*}{Observable} & \multicolumn{2}{c}{Data} && \multicolumn{1}{c}{Model} \\
		\cmidrule{2-5} 
		& Central value & $1\sigma$ range && Best fit \\
		\midrule
		$\theta_{12}^q$ $/^\circ$ &13.03 & 12.99 $\to$ 13.07 && 13.02  \\	
		$\theta_{13}^q$ $/^\circ$ &0.016 & 0.016 $\to$ 0.017 && 0.016  \\
		$\theta_{23}^q$ $/^\circ$ &0.189& 0.186 $\to$ 0.192 && 0.186 \\	
		$\delta^q$ $/^\circ$ & 69.22 & 66.12 $\to$ 72.31  && 70.66 \\
		$y_u$  $/ 10^{-6}$ & 3.060 & 2.111 $\to$ 4.009 && 3.253  \\	
		$y_c$  $/ 10^{-3}$ & 1.497 & 1.444 $\to$ 1.549 && 1.567  \\	
		$y_t$  			   & 0.666 & 0.637 $\to$ 0.694 && 0.611 \\
		$y_d$  $/ 10^{-4}$ & 1.473 & 1.311 $\to$ 1.635 && 1.614  \\	
		$y_s$  $/ 10^{-3}$ & 2.918 & 2.760 $\to$ 3.075 && 3.098 \\
		$y_b$   		   & 2.363 & 2.268 $\to$ 2.457 && 2.238  \\
		\bottomrule
	\end{tabular}
	\caption{Model predictions in the quark sector for  $\tan\beta = 20$, $M_{\mathrm{SUSY}} = 1$ TeV and $\bar{\eta}_b = -0.9$. The observables are at the GUT scale. The quark contribution to the total $\chi^2$ is 10.7.}
	\label{tab:numquarks}
\end{table}

\begin{figure}[ht]
	\centering
	\includegraphics[width=0.9\textwidth]{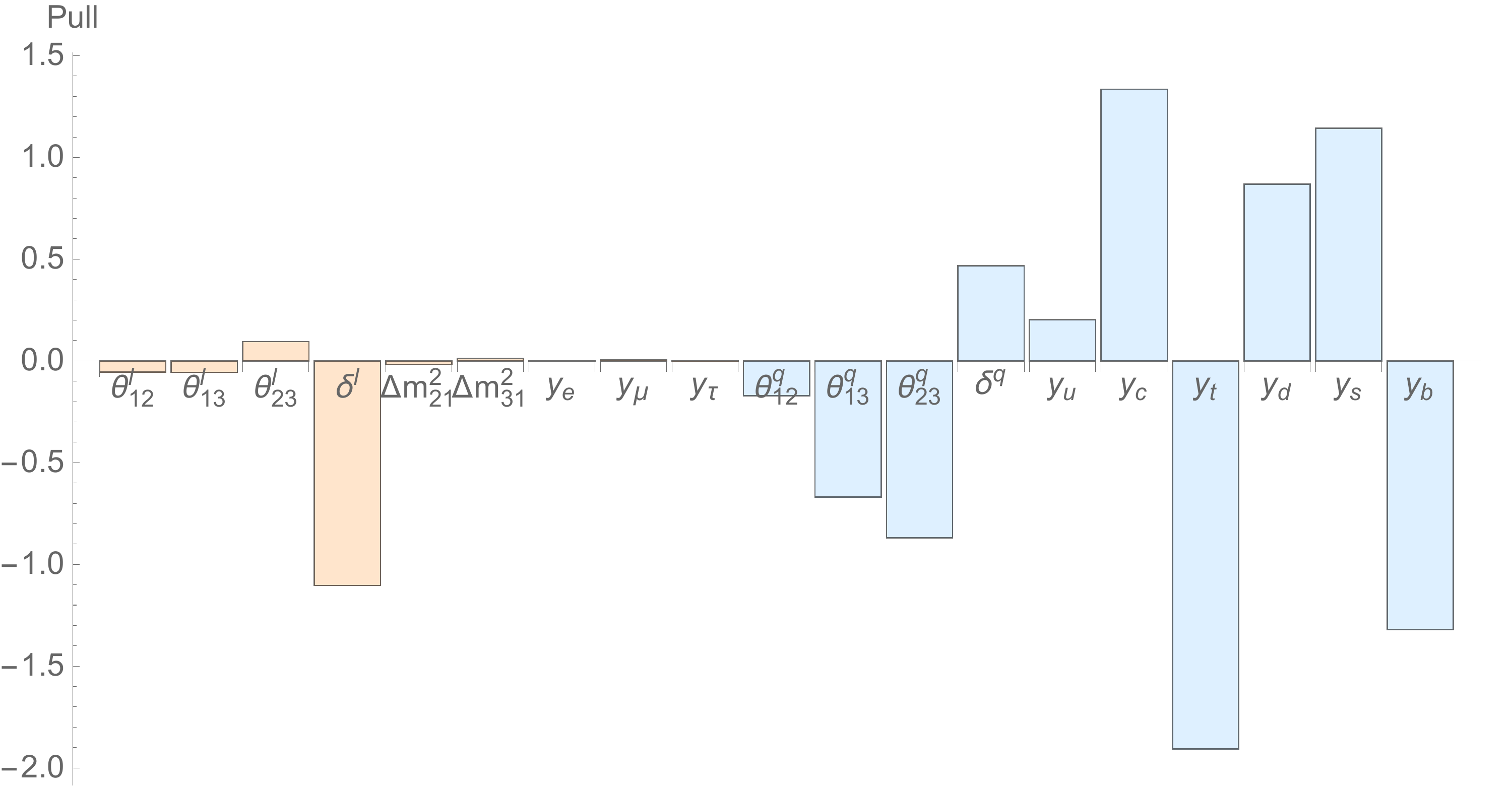}
	\caption{Pulls for the best fit of model to data, as shown in Tables \ref{tab:numleptons}-\ref{tab:numquarks}, for lepton (light orange) and quark (blue) parameters.}
	\label{fig:pulls}
\end{figure}

In table \ref{tab:numquarks}, we have all the quark Yukawa couplings and mixing parameters for the minimum $\chi^2$. The biggest contribution to the $\chi^2$ is coming from this sector, as shown in Fig.~\ref{fig:pulls}. This figure shows the corresponding pulls for lepton (light orange) and quark (blue) observables. As we can see, all parameters lie inside the $2\sigma$ region and the biggest pulls are in the quark Yukawa couplings. 

Table~\ref{tab:parameters} shows the input parameter values.\footnote{
Assuming the Dirac neutrino Yukawa parameters $y^\nu_i$ in Eq.~\ref{eq:natva}, we can compute the RHN masses, using the Seesaw formula in Eq.~\ref{eq:ssf} and taking the $\mu_i$ values from the fit, such that $M_1^R\sim 10^4$ GeV, $M_2^R\sim 10^{11}$ GeV and $M_3^R\sim 10^{15}$ GeV. Only $M_2$ has the expected natural value given in Eq.~\ref{eq:natva}. We remark that RHN Majorana masses come from unknown Planck suppressed physics, which is presumably responsible for the mismatch.} 
There are 13 real parameters plus two additional phases, a total of 15 input parameters to fit 19 data points. Naively, we can measure the goodness of the fit computing the reduced $\chi^2$, i.e. the $\chi^2$ per degree of freedom $\chi^2_\nu=\chi^2/\nu$. The number of degrees of freedom is given by $\nu=n-n_i$, where $n=19$ is the number of measured observables, while $n_i=15$ is the number of input parameters. A good fit is expected to have $\chi^2_\nu\sim1$. We have 4 degrees of freedom and the best fit has a reduced $\chi^2_\nu \simeq 3$. We view this as a good fit and it also remarks the predictivity of the model, not only fitting to all available quark and lepton data but also fixing the neutrino masses and Majorana phases. 

\begin{table}[ht]
	\centering
	\footnotesize
	\renewcommand{\arraystretch}{1.1}
	\begin{tabular}[t]{lr}
		\toprule
		Parameter & Value \\ 
		\midrule
		$y^u_1 \, /10^{-6}$ & 3.232 \\
		$y^u_2 \, /10^{-3}$ & 1.580 \\
		$y^u_3$ & $-0.610$ \\
		$y^d_{12} \, /10^{-4}$ &$-7.068$ \\
		$y^d_2 \, /10^{-4}$ & $-8.737$ \\
		$y^d_3$ & $-2.238$ \\
		\bottomrule
	
	\end{tabular}
	\hspace*{0.5cm}
	\begin{tabular}[t]{lr}
		\toprule
		Parameter & Value \\ 
		\midrule
		$y^e_{12}$ $/10^{-4}$ & $8.616$ \\
		$y^e_2$ $/10^{-2}$ & $1.013$ \\
		$y^e_3$  & $0.229$ \\
		$\mu_1 $ /meV & 6.845 \\
		$\mu_2 $ /meV & 27.18 \\
		$\mu_3 $ /meV & 42.17 \\
		\bottomrule
	\end{tabular}
	\hspace*{0.5cm}
	\begin{tabular}[t]{lr}
		\toprule
		Parameter & Value \\ 
		\midrule
		$y^P \, /10^{-4}$ & $2.475$ \\
		$\gamma$& $ 1.968\pi $ \\ 
		$\eta^\prime$ & $ 0.790\pi $ \\
		\bottomrule
	\end{tabular}
	\caption{Best fit input parameter values. The model has 13 real parameters: $y^u_i$, $y^d_i$, $y^e_i$, $\mu_i$ and $y^P$ and two additional free phases: $\eta^\prime$ and $\gamma$. The total $\chi^2$ is 11.9.} 
	\label{tab:parameters}
\end{table}

\subsection{SUSY threshold corrections}
\label{sec:susy}

The running of the MSSM Yukawa couplings to the GUT scale, $M_{GUT}$, was performed in \cite{Antusch:2013jca}. Here, the threshold corrections at the SUSY scale, $ M_\mathrm{SUSY} $, are parametrized by
\begin{equation}
\begin{split}
y^\mathrm{MSSM}_{u,c,t} 	&\simeq y^\mathrm{SM}_{u,c,t} \csc \bar{\beta} , \\
y^\mathrm{MSSM}_{d,s} 		&\simeq (1 + \bar{\eta}_q)^{-1} \, y^\mathrm{SM}_{d,s} \sec \bar{\beta} , \\
y^\mathrm{MSSM}_{b} 		&\simeq (1 + \bar{\eta}_b)^{-1} \, y^\mathrm{SM}_{b} \sec \bar{\beta} , \\
y^\mathrm{MSSM}_{e,\mu} 	&\simeq (1 + \bar{\eta}_\ell)^{-1} \, y^\mathrm{SM}_{e,\mu} \sec \bar{\beta} , \\
y^\mathrm{MSSM}_{\tau} 		&\simeq y^\mathrm{MSSM}_{\tau} \sec \bar{\beta} .
\end{split}
\label{eq:MSSMyukawas}
\end{equation}
The CKM parameters become
\begin{equation}
\begin{split}
\theta^{q,\mathrm{MSSM}}_{i3} 	\simeq \frac{1 + \bar{\eta}_b}{1 + \bar{\eta}_q} \, \theta^{q,\mathrm{SM}}_{i3} , \hspace{7mm}
\theta^{q,\mathrm{MSSM}}_{12} 	\simeq \theta^{q,\mathrm{SM}}_{12} , \hspace{7mm}
\delta^{q,\mathrm{MSSM}} 		\simeq \delta^{q,\mathrm{SM}} .
\end{split}
\label{eq:MSSMmixingangles}
\end{equation}
When running between $M_{SUSY}$ and $M_{GUT}$, the most relevant parameters are $\bar{\eta}_b$ and $\tan \bar{\beta}$. Due to their small contribtutions, we assume $ \bar{\eta}_q = \bar{\eta}_\ell = 0 $ and $\overbar{\beta}=\beta$. These assumptions do not affect the quality of the fit.
Similarly, we fix $ M_\mathrm{SUSY} = 1 $ TeV. The effect on the fit, of having it up to $ \mathcal{O}(10) $ TeV,  is minor.

Specifically, the parameter  $ \bar{\eta}_b$ is required to be somewhat large $ \bar{\eta}_b = -0.9 $ to obtain a good quality fit. Ignoring it would yield a fit of $\chi^2\sim 400$. The leading contributions to this parameter come from loops either sbottoms and gluinos or stops and higgsinos that add up to \cite{Hall:1993gn}
\begin{equation}
\bar{\eta}_b\simeq \frac{\tan\beta}{16\pi^2}\left(\frac{8}{3}g_3^2\frac{m_{\tilde{g}}\mu}{2m_0^2}+\lambda_t^2\frac{\mu A_t}{m_0^2}\right),
\end{equation}
where $m_0$ represents the squark masses, $g_3$ the strong coupling, $m_{\tilde{g}}$ the gluino mass and $A_t$ the SUSY softly breaking trilinear coupling involving the stops.
We see that a large contribution can be achieved when
\begin{equation}
m_{\tilde{g}},\mu,A_t >m_0,\hspace{7mm} \tan\beta \gtrsim 10.
\end{equation}
Since SUSY breaking lies beyond the scope of our paper, it is sufficient for us to show that there is a parameter space in the softly broken SUSY that generates the necessary corrections.

\subsection{Parameter counting}
\label{sec:parameter}
In this section we explain and clarify the number of parameters in our model. 
Clearly at the high energy scale there are many parameters associated with the undetermined $O(1)$
Yukawa couplings of the 43 superfields of the model. For example the renormalisable Yukawa superpotential in Eq.~\ref{eq:wy1} contains 23 parameters alone.
Then we must add to this all the $O(1)$ Yukawa couplings associated with vacuum alignment, GUT symmetry breaking and doublet-triplet splitting,
many of which we have not defined explicitly. Despite this, we are claiming that our model is predictive at low energies. How can this be?
The short answer is that most of these parameters are irrelevant for physics below the GUT scale, as discussed in detail below.

The effective fermion mass matrices generated below the GUT scale are given in Eq.~\ref{eq:masmat} as function of 18 free effective
parameters (remembering the constraint on $\eta$)
that will fix all the fermion masses and mixing angles, including RHN Majorana masses and Majorana phases. This compares favourably to the 31 parameters of a general high energy model, comprising 
21 parameters in the lepton sector of a general 3 right-handed neutrino seesaw model \cite{Davidson:2004wi}, plus the 6 quark masses and 4 CKM parameters.
However, below the seesaw scale of right-handed neutrino masses, the effective parameter counting is different again and requires further discussion below.

In order to perform the fit and compare our model with available data, we apply the seesaw mechanism to write the light effective left-handed Majorana neutrino mass matrix as a function of the new parameters $\mu_i$ in Eq~\ref{eq:ssf}. Therefore, we have 15 effective parameters at low energy (shown in Table~\ref{tab:parameters}) that fit the 19 so far measured or constrained observables in Fig.~\ref{fig:pulls}
.\footnote{ We need to run up to the GUT scale these observables and, therefore, we need to include SUSY threshold corrections. The fit is
therefore also dependent on 
$\overbar{\eta}_b$ and $\tan \beta$. As shown earlier, we find a good fit for $\overbar{\eta}_b=-0.9$ and $\tan \beta=20$.
} After the fit is performed, the model predicts all the three light neutrino masses with a normal ordering, 
a CP oscillation phase of $260^\circ$ and both the Majorana phases, corresponding to a total of 22 low energy observables
which will be eventually observable
(10 from the quark sector discussed above and 12 from the lepton sector, including the two Majorana phases).
Therefore we see that, below the seesaw scales, the model contains 15 effective parameters which generate 22 observables,
making the model eminently testable, as these observables become better determined.

\section{ \boldmath{$N_2$} Leptogenesis}
\label{sec:lept}
The source of the Baryon Asymmetry of the Universe (BAU)
\begin{equation}
\eta_{B}^{CMB}=(6.1\pm 0.1)\times 10^{-10},
\end{equation}
remains unexplained in the SM. One of the most convincing sources for it is Leptogenesis, where the asymmetry is generated through CP breaking decays of heavy RHNs into leptons, then converted into baryons through sphalerons \cite{DiBari:2012fz}.

The simplest mechanism to generate the correct BAU, happens when the lightest RHNs has CP breaking decays and has a mass of about $\sim 10^{10} GeV$. In our model, according to Eq. \ref{eq:natva}, it is the second RHN the one that is expected to be at that scale. When leptogenesis is generated mainly by the decays of the second RHN it is called $N_2$ leptogenesis. This has already been calculated in \cite{DiBari:2015svd} and we will apply such calculations to our specific model.

\subsection{General  \boldmath{$N_2$} leptogenesis.}
Leptogenesis calculations  are done in the so called Flavor Basis, where the charged lepton and RHN mass matrices are diagonal and we work with the Dirac neutrino mass matrix
\begin{equation}
\begin{split}
m_{D}&=V_{eL}M_D^\nu U_N^T,\\
V_{eL}M^{e\dagger}M^eV_{eL}^\dagger=diag(y_e^2,y_\mu^2,y_\tau^2)v_d^2,&\hspace{7mm}
U_N M^R U_N^T=diag(M_1,M_2,M_3).
\label{eq:flaba}
\end{split}
\end{equation}
The total and flavoured decay parameters, $K_i$ and $K_{i\alpha}$ respectively, can be written as
\begin{equation}
K_{i\alpha}  = {|m_{D\alpha i}|^2 \over m_{\star}^{MSSM}\, M_i}  
\hspace{5mm}\mbox{\rm and} \hspace{5mm}
K_i = \sum_{\alpha}\,K_{i\alpha}= {(m^{\dagger}_D \, m_D)_{ii} \over m_{\star}^{MSSM}\, M_i} \,  ,
\end{equation}
where the  equilibrium neutrino mass is given by 
\begin{equation}\label{mstar}
m_{\star}^{MSSM}  \simeq 0.78\times 10^{-3}\,{\rm eV} \, \sin^2\beta.
\end{equation}
The wash-out at the production 
is described by the efficiency factor $\kappa(K_{2\alpha})$ that for an initial thermal 
$N_2$ abundance can be calculated as
\begin{equation}\label{kappa}
\kappa(K_{2\alpha}) = 
{2\over z_B(K_{2\alpha})\,K_{2\alpha}}
\left(1-e^{-{K_{2\alpha}\,z_B(K_{2\alpha})\over 2}}\right) \,  , \;\; z_B(K_{2\alpha}) \simeq 
2+4\,K_{2\alpha}^{0.13}\,e^{-{2.5\over K_{2\alpha}}} \,   .
\end{equation}

In the hierarchical RH neutrino mass limit, as our model is, the CP asymmetries can be approximated to
\begin{equation} \label{eps2al}
\varepsilon_2=\sum_\alpha \varepsilon_{2\alpha},\ \ \varepsilon_{2\alpha} \simeq  {3\over 8\,\pi}\frac{\,M_2\,}{v^2}\, {{\rm Im} \Big[ \big(m_D^\dagger \big)_{i \alpha} \, \big(m_D \big)_{\alpha 3}
\big(m_D^\dagger m_D \big)_{i3} \Big]\over M_2\,M_3\,\widetilde{m}_2 \, } ,
\end{equation}
where $\widetilde{m}_2 \equiv (m_D^{\dag}\, m_D)_{2 2}/M_2$.

In the regime where  $5\times 10^{11}\,{\rm GeV}\,(1+\tan^2\beta) \gg M_2 \gg 5\times 10^{8}\,{\rm GeV}\,(1+\tan^2\beta)$,  the  final $B-L$ asymmetry can  be calculated using
\begin{eqnarray}\label{twofl} \nonumber
N_{B-L}^{\rm f} & \simeq &
\left[{K_{2e}\over K_{2\tau_2^{\bot}}}\,\varepsilon_{2 \tau_2^{\bot}}\kappa(K_{2 \tau_2^{\bot}}) 
+ \left(\varepsilon_{2e} - {K_{2e}\over K_{2\tau_2^{\bot}}}\, \varepsilon_{2 \tau_2^{\bot}} \right)\,\kappa(K_{2 \tau_2^{\bot}}/2)\right]\,
\, e^{-{3\pi\over 8}\,K_{1 e}}+ \\ \nonumber
& + &\left[{K_{2\mu}\over K_{2 \tau_2^{\bot}}}\,
\varepsilon_{2 \tau_2^{\bot}}\,\kappa(K_{2 \tau_2^{\bot}}) +
\left(\varepsilon_{2\mu} - {K_{2\mu}\over K_{2\tau_2^{\bot}}}\, \varepsilon_{2 \tau_2^{\bot}} \right)\,
\kappa(K_{2 \tau_2^{\bot}}/2) \right]
\, e^{-{3\pi\over 8}\,K_{1 \mu}}+ \\
& + &\varepsilon_{2 \tau}\,\kappa(K_{2 \tau})\,e^{-{3\pi\over 8}\,K_{1 \tau}} \,  ,
\label{eq:nbl2f}
\end{eqnarray}
where we indicated with $\tau_2^{\bot}$ the electron plus muon component of the 
quantum flavour states produced by the $N_2$-decays defining 
$K_{2\tau_2^{\bot}}\equiv K_{2e}+K_{2\mu}$, $\varepsilon_{2\tau_2^{\bot}}\equiv \varepsilon_{2e}+\varepsilon_{2\mu}$.
The final asymmetry, in terms of the baryon to photon number ratio is
\begin{equation}
\eta_B\simeq 2\ a_{sph}\frac{N_{B-L}}{N_\gamma^{rec}},
\end{equation}
where $\alpha_{sph}=8/23$ is the fraction of $B-L$ asymmetry converted into baryon asymmetry by sphalerons. The photon asymmetry at recombination is $(N^{rec}_\gamma)^{MSSM}\simeq 78$. The factor of 2 accounts for the asymmetry generated by the RH neutrinos and sneutrinos. 

\subsection{Leptogenesis in our model}

Using the matrices in Eq.~\ref{eq:masmat} and the fit in Table~\ref{tab:parameters}, we may calculate the BAU generated through $N_2$ Leptogenesis in our model. The first thing to note is that the parameters are quite hierarchical so that the rotation angles of the diagonalizing matrices can be neglected since they only give $1\%$ contributions
\begin{equation}
V_{eL}\simeq \mathbb{1},\hspace{7mm} U_N\simeq diag(e^{-i\eta/2},0,e^{-i\eta'/2}), 
\end{equation}
and the neutrino mass matrix in the Flavor Basis becomes
\begin{equation}
m_{Dij}\simeq\left(\begin{array}{ccc}
y_1^\nu  e^{i\eta/2}& 2y_1^\nu e^{i\eta} & 0
\\2\  y_1^\nu  e^{i\eta/2}& y_2^\nu & y_2^\nu e^{-i\eta'/2}
\\ 0 & y_2^\nu & y_3^\nu e^{i\eta'/2}
\end{array}\right)v_u.
\label{eq:diracmassFB}
\end{equation}
Also, due to the hierarchical nature of the couplings we may safely assume that the RHN mass parameter are equal to their mass eigenvalues $M^R_a\simeq M_a$.

One of the features of the matrix structure is that $K_{1\tau}$ vanishes,
due to the approximate zero in the (3,1) entry of the Dirac mass matrix \footnote{The zero is a consequence of the CSD2 vaccum alignment; it would not be zero for CSD3 vacuum alignment.} , so that 
the last term in Eq. \ref{eq:nbl2f} is greatly enhanced since it overcomes the exponential suppression. With these approximations, the baryon asymmetry becomes
\begin{equation}
\begin{split}
\eta_B&\simeq \frac{2\alpha_{sph}}{N^{rec}_\gamma}\ \kappa(K_{2\tau})\ \varepsilon_{2\tau},\\
K_{2\tau}=\frac{(y_2^\nu)^2v_u^2}{m_\star^{MSSM}M_2}, &\hspace{7mm}\epsilon_{2\tau}=\sin\eta'\frac{3}{8\pi}\frac{M_2 }{M_3 }\frac{(y_3^\nu)^2}{2}\sin^2\beta.
\end{split}
\end{equation}
We note that $\eta'$ is identified with the leptogenesis phase. With use of Eq. \ref{eq:ssf}, we may write the neutrino Yukawa couplings as
 $y^\nu_a=\sqrt{\mu_a M^R_a}/v_u$  so that
\begin{equation}
\eta_B\simeq\sin\eta'\frac{3}{8\pi}\frac{\alpha_{sph}}{N^{rec}_\gamma}\ \kappa\left(\frac{\mu_2}{m_\star^{MSSM}}\right) \frac{\mu_3 M_2}{v^2},
\end{equation}
where we note that the only free parameter is $M_2$. Using the parameters from the fit, in Table \ref{tab:parameters}, the correct BAU is generated when
\footnote{
$M_2$ has been computed numerically, including the rotation angles of the diagonalizing matrices in Eq.~\ref{eq:diracmassFB}. }
\begin{equation}
M_2\simeq 1.9\times 10^{11}\ GeV.
\end{equation}
From Eq. \ref{eq:natva} we see that this is the natural value for the second RHN mass, so that the model naturally explains the origin of the BAU through $N_2$ leptogenesis without any need for tuning.

\section{Conclusion}
\label{sec:con}
We have constructed a SUSY GUT of flavour based on the symmetry $S_4\times SO(10)\times \mathbb{Z}_4^3\times \mathbb{Z}_4^R$ that is relatively simple, predictive and fairly complete.
The Higgs sector of the model involves two $SO(10)$ 10-plets, a 16-plet and its conjugate representation, 
and three 45-plets. These low dimensional Higgs representations are all that is required to break the GUT symmetry,
yield the Clebsch relations responsible for the difference of the charged fermion masses, and account for 
heavy Majorana right-handed neutrino masses.
In order to account for the hierarchical mixing structure of the Yukawa matrices,
we also need a particular set of $S_4$ triplet flavons with hierarchical VEVs and particular CSD2 vacuum alignments,
where both features are fully discussed here.
To complete the model we also require a rather rich spectrum of high energy 
messenger and alignment superfields, which, like most of the Higgs fields, do not appear in the low energy effective
theory. 

We highlight and summarise the main successes and features of the model as follows:
\begin{itemize}
\item The model is succesfully built with an $SO(10)$ gauge symmetry where all of the fields belong to the small ``named" representations: fundamental, spinorial and adjoints; this could be helpful for a possible future string embedding.
\item It contains a superpotential that spontaneously breaks the original symmetry:
\\ $S_4\times SO(10)\times \mathbb{Z}_4^3\times \mathbb{Z}_4^R\to SU(3)_C\times SU(2)_L\times U(1)_Y\times \mathbb{Z}_2^R$. The model also spontaneously breaks $CP$.
\item The $S_4$ breaking superpotential that yields the CSD2 vacuum alignment is fairly simple.
\item All the GUT scale parameters are natural and $\sim O(1)$, explaining the hierarchy of the low energy parameters,
where the family mass hierarchy is due the derived hierarchy of flavon VEVs $|v_1|\ll |v_2| \ll |v_3|$,
rather than by Froggatt-Nielsen.
\item The model contains a working doublet-triplet mechanism, that yields exactly two light Higgs doublets 
from two $SO(10)$ Higgs multiplets, respectively and without mixing, apart from the $\mu$ term which is generated at the correct scale. It also has well behaved proton decay. 
%\item Even though at high energies, the model contains more than two doublets, only two of them are light. In fact all the fields, except the MSSM ones, obtain a mass at the GUT scale, ensuring Gauge Coupling Unification at the GUT scale.
\item The model naturally generates sufficient BAU  through $N_2$ Leptogenesis,
which fixes the second right-handed neutrino mass 
$M_2\simeq 2\times 10^{11}$ GeV,
in the natural range predicted by the model.
\item At low energies, the model contains 15 free parameters that generate 19  presently constrained observables so that it is quite predictive.
The model achieves an excellent fit of the SM fermion masses and mixing angles, with $\chi^2=11.9$.
\end{itemize}

We find it remarkable that all of the above can be achieved consistently within a single model.
It contains 43 supermultiplet fields, which is the minimal number for any such complete model
in the literature so far.

Despite the above successes of the model, it also has a few drawbacks. It does not explain SUSY breaking, and it relies on specific threshold corrections. Even though it has an almost complete UV completion, it still relies on $O(1/M_P)$ terms 
for the right-handed neutrino masses. Indeed $M_1$ and $M_3$ apparently
do not have such natural values as $M_2$, and we are forced to explain this away by 
appealing to the unknown physics at the Planck scale. 
The symmetry breaking superpotential gives VEVs to most of the GUT breaking fields but it does not drive all of them.
Also we do not address the strong CP problem, inflation or Dark Matter (DM) (which may in principle be due to the 
lightest SUSY particle, stabilised by the R-parity). Indeed we have not considered the SUSY spectrum at all.
Such issues are beyond the stated aims of the present paper, 
which is to propose a complete grand unified theory of flavour and leptogenesis, consistent with the latest 
data on quark and lepton masses and mixing parameters,
in which the three families of quarks and leptons are unified into a single $(3',16)$ representation of $S_4\times SO(10)$.

Importantly, the model can be tested due to its robust predictions of 
a normal neutrino mass ordering, 
a CP oscillation phase
of $260^{\circ}$, an atmospheric angle of $42^{\circ}$ in the first octant and a neutrinoless double beta 
decay parameter $m_{\beta \beta}= 11$ meV, with the sum of neutrino masses being 76 meV. 
These predictions, together with the other lepton 
mixing angles given earlier, will enable the model to be tested by the forthcoming neutrino experiments.

\subsection*{Acknowledgements}
We thank Fredrik Bj\"{o}rkeroth for discussions.
S.\,F.\,K. acknowledges the STFC Consolidated Grant ST/L000296/1.
This project has received funding from the European Union's Horizon 2020 research and innovation programme under the Marie Sk\l{}odowska-Curie grant agreements 
Elusives ITN No.\ 674896 and
InvisiblesPlus RISE No.\ 690575.

\end{document}